\documentclass[10pt,journal]{IEEEtran}
\selectfont

\usepackage[utf8]{inputenc}
\usepackage{multirow}
\usepackage{acronym}
\usepackage{amssymb}
\usepackage{pifont}
\usepackage{rotating}
\usepackage{url}
\usepackage{soul}
\usepackage{array}
\usepackage{amsmath}
\usepackage{makecell}
\usepackage{breakurl}
\usepackage{balance}
\usepackage{xcolor}
\usepackage{hyperref}
\usepackage{xr}
\usepackage[nocompress]{cite}
\usepackage{verbatim}

\acrodef{EH}{Energy Harvesting}
\acrodef{RF}{Radio Frequency}
\acrodef{WPT}{Wireless Power Transfer}
\acrodef{WPCN}{Wireless Powered Communication Network}
\acrodef{SWIPT}{Simultaneous Wireless Information and Power Transfer}
\acrodef{SINR}{Signal to Noise plus Interference Ratio}
\acrodef{SISO}{Single-Input Single-Output}
\acrodef{MISO}{Multiple-Input Single-Output}
\acrodef{MIMO}{Multiple-Input Multiple-Output}
\acrodef{mMIMO}{Massive \ac{MIMO}}
\acrodef{SIMO}{Single-Input Multiple-Output}
\acrodef{DF}{Decode and Forward}
\acrodef{AF}{Amplify and Forward}
\acrodef{AWGN}{Additive White Gaussian Noise}
\acrodef{CSI}{Channel State Information}
\acrodef{PS}{Power Splitting}
\acrodef{TS}{Time Splitting}
\acrodef{SR}{Separate Receiver}
\acrodef{IoT}{Internet of Things}
\acrodef{WSN}{Wireless Sensor Network}
\acrodef{SOP}{Secrecy Outage Probability}
\acrodef{SRM}{Secrecy Rate Maximization}
\acrodef{DoS}{Denial of Service}
\acrodef{VLC}{Visible Light Communications}
\acrodef{CRN}{Cognitive Radio Network}
\acrodef{OFDM}{Orthogonal Frequency Division Multiplexing}
\acrodef{NOMA}{Non-Orthogonal Multiple Access}
\acrodef{RSMA}{Rate-Splitting Multiple Access}
\acrodef{SEE}{Secrecy Energy Efficiency}
\acrodef{WPTN}{Wireless Power Transfer Network}
\acrodef{FDMA}{Frequency Division Multiple Access}
\acrodef{TDMA}{Time Division Multiple Access}
\acrodef{OFDMA}{Orthogonal Frequency Division Multiplexing Access}
\acrodef{SDR}{Software Defined Radio}
\acrodef{ESR}{Ergodic Secrecy Rate}
\acrodef{UAV}{Unmanned Aerial Vehicles}
\acrodef{HPPP}{Homogeneous Poisson Point Process}
\acrodef{QoS}{Quality of Service}
\acrodef{ICA}{Independent Component Analysis}
\acrodef{CP-ABE}{Ciphertext Policy - Attribute Based Encryption}
\acrodef{AES}{Advanced Encryption Standard}
\acrodef{ECDSA}{Elliptic Curve Digital Signature Algorithm}
\acrodef{SDN}{Software Defined Networking}
\acrodef{PDR}{Packet Delivery Ratio}
\acrodef{WBAN}{Wireless Body Area Network}
\acrodef{D2D}{Device-to-Device Communications}
\acrodef{RFID}{Radio Frequency IDentification}
\acrodef{NFC}{Near Field Communication}
\acrodef{ABCN}{Ambient Backscatter Communication Network}
\acrodef{CPS}{Cyber-Physical Systems}
\acrodef{NIC}{Network Interface Card}
\acrodef{IRS}{Intelligent Reconfigurable Surfaces}
\acrodef{LIS}{Large Intelligent Surfaces}

\title{Security in Energy Harvesting Networks:\\ A Survey of Current Solutions and Research Challenges}
\author{Pietro Tedeschi, Savio Sciancalepore, and Roberto Di Pietro
\\Division of Information and Computing Technology (ICT) \\ College of Science and Engineering (CSE), Hamad Bin Khalifa University (HBKU) \\ Doha, Qatar \\
  e-mail: \{ptedeschi, ssciancalepore, rdipietro\}@hbku.edu.qa}

\newcolumntype{P}[1]{>{\centering\arraybackslash}p{#1}}
\newcommand{\cmark}{\ding{51}}%
\newcommand{\xmark}{\ding{55}}%
\newcommand{\xmarklight}{\ding{53}}%

\newcommand\blfootnote[1]{%
  \begingroup
  \renewcommand\thefootnote{}\footnote{#1}%
  \addtocounter{footnote}{-1}%
  \endgroup
}

\graphicspath{{figs/}}

\begin{document}
\bstctlcite{IEEEexample:BSTcontrol}

\IEEEtitleabstractindextext{%
\begin{abstract}
\blfootnote{This is a personal copy of the authors. Not for redistribution. The final version of the paper is available through the IEEExplore Digital Library, at the link: \url{XXX}, with the DOI: \url{XXX.}}
The recent advancements in hardware miniaturization capabilities have boosted the diffusion of systems based on Energy Harvesting (EH) technologies, as a means to power embedded wireless devices in a sustainable and low-cost fashion. Despite the undeniable management advantages, the intermittent availability of the energy source and the limited power supply has led to challenging system trade-offs, resulting in an increased attack surface and a general relaxation of the available security services.

In this paper, we survey the security issues, applications, techniques, and challenges arising in wireless networks powered via EH technologies. We explore the vulnerabilities of EH networks, and we provide a comprehensive overview of the scientific literature, including attack vectors, cryptography techniques, physical-layer security schemes for data secrecy, and additional physical-layer countermeasures. For each of the identified macro-areas, we compare the scientific contributions across a range of shared features, indicating the pros and cons of the described techniques, the research challenges, and a few future directions. 
Finally, we also provide an overview of the emerging topics in the area, such as Non-Orthogonal Multiple Access (NOMA) and Rate-Splitting Multiple Access (RSMA) schemes, and Intelligent Reconfigurable Surfaces, that could trigger the interest of industry and academia and unleash the full potential of pervasive EH wireless networks.

\end{abstract}

}

\maketitle

\IEEEdisplaynontitleabstractindextext

%
\IEEEpeerreviewmaketitle


\section{Introduction}
\label{sec:intro}

\textcolor{black}{Next-generation wireless networks are increasingly evolving toward pervasive and ubiquitous systems, able to bring new capabilities in a variety of environments and application scenarios~\cite{Agiwal2016_ieeest}. This paradigm evolution has been enabled in the late 1990s thanks to the rise of \acp{WSN}, and it has progressed to a large scale thanks to the evolution of \ac{IoT}, \ac{CPS}, and, more recently, 5G networks~\cite{Akpakwu2018_access}.}

Despite the massive growth and the increasing computational capabilities of tiny wireless devices, energy availability and consumption still remain critical issues~\cite{Alsaba2018_ieeest}. Indeed, while improvements in hardware embedding capabilities widened the computing efficiency and potential applications of wireless devices, the energy budget required to power the devices increases accordingly, severely affecting the lifetime and availability of the associated wireless networks. At the same time, increasing the capacity of batteries on-board of these devices is only a time-limited solution, further postponing the issue and impacting on the form-factor and ubiquity of the devices~\cite{Ulukus2015_jsac}.

\textcolor{black}{These issues motivated the recent exponential growth of the \emph{\ac{EH}} research area. Several solutions, such as solar, thermoelectric, and mechanical harvesting techniques, have been re-introduced and improved in their size and features to be suitable to power embedded devices, in a tentative to move toward perpetual battery-free pervasive communications~\cite{He2015_comma}. At the same time, the theoretical studies describing the usage of existing or dedicated \ac{RF} transmissions to power embedded devices have finally translated into real devices and hit the market, thanks to dedicated prototypes and commercial products building on RF power harvesting to provide energy to embedded devices \cite{Guo2016_comma}. The potentials of EH for the next-generation low-power wireless networks are confirmed by several projects, funded by the most important funding agencies throughout the world. For instance, the EU funded several projects related to EH under the Horizon 2020 (H2020) research and innovation program, including Tryst Energy~\cite{tryst}, DuraCap~\cite{duracap}, EnABLES~\cite{enables}, Plug-N-Harvest~\cite{plug-n-harvest}, and Re-Vibe~\cite{re-vibe}, to name a few. The US National Science Foundation also funded several pilot studies, such as CORE~\cite{core_nsf}, SMALL~\cite{small_nsf}, and CAREER~\cite{career_nsf}. In addition, the National Science Foundation of China is also supporting several scientific contributions focusing on the development of EH technologies, such as~\cite{wang2020_tc},~\cite{dong2020_iotj}, and~\cite{ye2020_iotj}.}

This is finally confirmed also by recent pilot studies, indicating EH techniques such as ambient backscatter and \ac{RF} energy harvesting as crucial enabling technologies for \emph{sixth generation} (6G) wireless networks \cite{Giordani2019, Saad2019, Rappaport2019}.

\begin{figure*}[htbp]
    \centering
    \includegraphics[width=\textwidth]{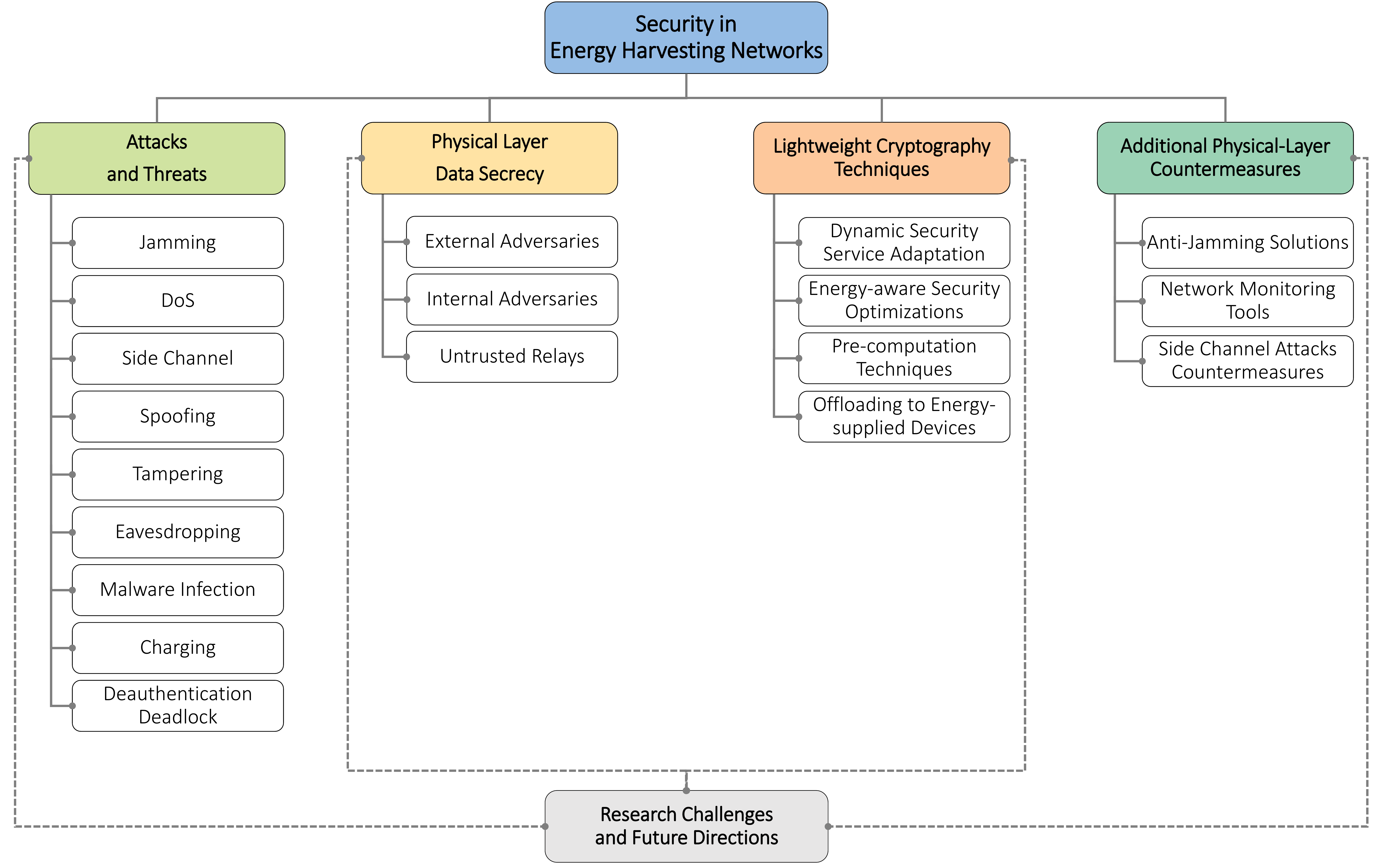}
    \caption{Taxonomy and classification of the scientific contributions in this survey. We identified four macro-areas related to security issues in EH networks, being them dedicated to the investigation of attacks, physical-layer data secrecy issues, lightweight cryptography approaches, and additional physical-layer countermeasures. Within each area, we further identify distinguishing adversarial models and specific design trends. Finally, for each of the investigated topics, we identify crucial research challenges and future directions .}
    \label{fig:taxonomy}
\end{figure*}

\textcolor{black}{Despite their appealing features and significant advantages, the switch to EH sources generally widens the attack surface of wireless networks and cyber-physical systems. On the one hand, powering embedded devices via intermittent energy sources and, in the case of wireless powering, removing wired power connections, expose the devices to inexpensive and easy eavesdropping attacks~\cite{perera2018_comst},~\cite{liu2016_commag},~\cite{vo2017}. On the other hand, the shift to EH power sources exacerbates the well-known trade-off between energy and security, further decreasing the energy budget available for security operations~\cite{khan2019_comst}. Considering the traditional low awareness of security vulnerabilities by wireless network administrators (for the sake of immediate services availability), the reduction of the energy budget often brings to the reduction of security barriers~\cite{Butun2019_ieeest}.}

\textcolor{black}{Security issues in \ac{EH} networks have been quickly touched by few position papers and surveys available in the literature. An overview of their main features and differences with this contribution is provided in Tab. \ref{tab:related_detail}.}

\begin{table*}[htbp]
\centering
\color{black}
\caption{Comparison between the topics tackled in this paper and the related surveys touching network security issues in EH networks.}
\begin{tabular}{|P{0.8cm}|P{1.1cm}|P{1.1cm}|P{1.5cm}|P{1.1cm}|P{1.8cm}|P{1.6cm}|P{1.5cm}|P{1.5cm}|}
\hline
\textbf{Ref.} & \textbf{RF Sources} & \textbf{Solar Sources} & \textbf{Mechanical Sources} & \textbf{Wind Sources} & \textbf{Thermoelectric Sources} & \textbf{Cryptography Approaches} & \textbf{Additional PHY-layer countermeasures} & \textbf{Research Challenges} \\
\hline \hline
\cite{dimauro2012_peccs} & \checkmark & \xmarklight & \xmarklight & \xmarklight & \xmarklight & \xmarklight & \xmarklight & \xmarklight \\ \hline
\cite{kang2015_commag} & \checkmark & \xmarklight & \xmarklight & \xmarklight & \xmarklight & \xmarklight & \xmarklight & \xmarklight \\ \hline
\cite{liu2016_commag} & \checkmark & \xmarklight & \xmarklight & \xmarklight & \xmarklight & \xmarklight & \xmarklight & \xmarklight \\ \hline
\cite{chen2016_wc} & \checkmark & \xmarklight & \xmarklight & \xmarklight & \xmarklight & \xmarklight & \xmarklight & \xmarklight \\ \hline
\cite{zhao2017_access} & \checkmark & \xmarklight & \xmarklight & \xmarklight & \xmarklight & \xmarklight & \xmarklight & \xmarklight \\ \hline
\cite{chen2018_wc} & \checkmark & \xmarklight & \xmarklight & \xmarklight & \xmarklight & \xmarklight & \xmarklight & \checkmark \\ \hline
\cite{wang2018_commag} & \checkmark & \xmarklight & \xmarklight & \xmarklight & \xmarklight & \xmarklight & \xmarklight & \xmarklight \\ \hline
\cite{perera2018_comst} & \checkmark & \xmarklight & \xmarklight & \xmarklight & \xmarklight & \xmarklight & \xmarklight & \xmarklight \\ \hline
\cite{alsaba2018_comst} & \checkmark & \xmarklight & \xmarklight & \xmarklight & \xmarklight & \xmarklight & \xmarklight & \xmarklight \\ \hline
\cite{jindal2019_springer} & \checkmark & \xmarklight & \xmarklight & \xmarklight & \xmarklight & \xmarklight & \xmarklight & \xmarklight \\ \hline
\cite{hossain2019_access} & \checkmark & \xmarklight & \xmarklight & \xmarklight & \xmarklight & \xmarklight & \xmarklight & \checkmark \\ \hline
\cite{jameel2019} & \checkmark & \xmarklight & \xmarklight & \xmarklight & \xmarklight & \xmarklight & \checkmark & \checkmark \\ \hline
\cite{hamamreh2019_comst} & \checkmark  & \xmarklight & \xmarklight & \xmarklight & \xmarklight & \xmarklight & \xmarklight & \xmarklight \\ \hline
\textbf{This survey} & \cmark & \cmark & \cmark & \cmark & \cmark & \cmark & \cmark & \cmark \\ \hline
\end{tabular}
\label{tab:related_detail}
\end{table*}
\color{black}

\textcolor{black}{
Several pioneering contributions, such as \cite{dimauro2012_peccs}, \cite{kang2015_commag}, \cite{liu2016_commag}, and \cite{chen2016_wc}, preliminarily identified the main attacks and security issues affecting RF EH networks, inspiring a relevant literature production over the last years. However, besides considering only a limited set of weaknesses in wireless-powered EH networks, the cited contributions did not provide any overview of possible defense schemes and countermeasures. Similarly, the position paper by the authors in \cite{chen2018_wc} described the usage of cooperative strategies in wireless-powered EH networks, highlighting their benefits also from the security perspective. However, the cited paper did not consider the literature in the security domain, and security issues were not addressed, too. Looking at the main surveys available in the literature, valuable contributions are provided by the authors in \cite{perera2018_comst}, \cite{zhao2017_access}, \cite{alsaba2018_comst}, and \cite{hossain2019_access}. However, they focused either on the description of the main aspects and challenges of RF EH technologies (such as \cite{perera2018_comst}), or on specific techniques and phenomena, such as interference \cite{zhao2017_access}, beamforming \cite{alsaba2018_comst}, and \ac{SWIPT} technologies \cite{hossain2019_access}. We also report the recent survey by the authors in~\cite{hamamreh2019_comst} on physical-layer security, where a few physical-layer security approaches based on EH concepts are mentioned, without any further detail. Finally, the recent study by the authors in \cite{jindal2019_springer} only provides an overview of the techniques used to improve physical layer security in SWIPT networks. However, the main attacks, defensive technologies, network availability considerations, countermeasures, and research challenges related to the security domain are not discussed. In addition, we notice that none of the surveys and position papers currently available in the literature widens the analysis, practically excluding other EH sources, such as solar, wind, mechanical, and thermoelectric sources that, instead, we take into full consideration in this survey. Further, the cited contributions  also did not consider all the works that adapted cryptography approaches to work efficiently on EH nodes.
\\
\noindent
Therefore, we remark that the literature is actually missing a comprehensive survey specifically conceived to tackle all the security aspects connected to the operation of EH networks, including information secrecy at the physical layer, cryptography techniques, attacks, additional physical-layer defensive technologies, and, finally, network availability considerations. }

{\bf Contribution.} In this paper, we close the above-introduced gap by providing a comprehensive survey on security issues, applications, approaches, and research challenges characterizing wireless networks and embedded systems powered via \acl{EH} sources. Building on the existing literature, we first describe the threats and vulnerabilities affecting wireless networks operating thanks to EH sources. Next, we survey the security schemes related to EH networks available in the literature, classifying them across reference macro-areas, such as physical-layer approaches for data secrecy, lightweight cryptography techniques, and additional physical-layer countermeasures. For each of these macro-categories, as a novel contribution, we describe the adversarial models, the assumptions, the solutions, and we cross-compare the scientific works available in the literature via their distinctive features (see Figure~\ref{fig:taxonomy} for a high-level overview).

Finally, we provide future research directions along the above-described macro-areas, as well as, some additional crucial research challenges, whose further investigation could unleash the full potential of EH technologies, toward the large-scale adoption of EH sources in wireless networks.

\textcolor{black}{We remark that this paper is not yet another survey on physical-layer security. Indeed, in this paper we survey security issues, approaches, and applications strictly connected to all \ac{EH} networks (not only RF), where dedicated RF and natural energy sources are used to enable \ac{D2D} communications in IoT and generic wireless networks. 
These networks include embedded systems powered via solar, mechanical, wind, and electromagnetic sources, such as ambient backscatter, \acl{WPCN}, \ac{SWIPT} networks, to name a few. Therefore, technologies such as \ac{RFID}, \ac{NFC}, and traditional tag-to-reader backscatter networks are not within the scope of this survey, as these communication schemes can neither power embedded systems, nor detect transmissions of other nearby devices, nor power D2D networks, but they can communicate exclusively to compatible devices (e.g. RFID and NFC readers)~\cite{Liu2013} ---the interested readers can refer to~\cite{juels2006} and~\cite{coskun2013survey} for a comprehensive survey on the security issues related to the cited technologies.}

\textcolor{black}{
{\bf Roadmap.} The rest of this paper is organized as follows. Section~\ref{sec:techno} provides an overview of the technologies enabling energy harvested embedded systems; then, Section~\ref{sec:attacks} describes the threats and vulnerabilities intrinsic to the operation of EH networks, motivating the study of several security schemes in the areas.
The application and customization of cryptography schemes in EH networks is discussed in Section~\ref{sec:light_crypto}. Then, the mechanisms and strategies deployed to enhance physical-layer data secrecy are described in Section~\ref{sec:phy_secrecy}. Additional physical-layer countermeasures deployed to face the afore-mentioned attacks are described in Section~\ref{sec:defense}. Emerging research areas and future research directions are described in Section~\ref{sec:challenge} and, finally, Section~\ref{sec:conclusion} tightens conclusions.
}

\section{Overview of Enabling Technologies}
\label{sec:techno}

In this section, we introduce the enabling technologies considered in this work, as well as some important background concepts that have been introduced in the literature when dealing with security in EH networks. Specifically, Section~\ref{sec:ext} reports some basic notions on energy harvesting technologies using non-RF sources, such as \textcolor{black}{solar-based sources}, mechanical sources, and thermoelectric generators, while RF harvesting methods are introduced in Section~\ref{sec:rf_charge}. Preliminary details regarding \ac{ABCN} are introduced in Section~\ref{sec:back}, while Section~\ref{sec:wpcn} focuses on the principles enabling \aclp{WPCN}. Section~\ref{sec:swipt} provides an overview of the logic and hardware architectures adopted in \ac{SWIPT} networks. Finally, few distinctive features of EH networks crucial for the design of security solutions are discussed in Section~\ref{sec:eh_chal}. \textcolor{black}{We highlight that the aim of this section is to let the reader familiarize with harvesting technologies and their main features, with reference to security issues. For exhaustive discussion on the main hardware considerations behind EH technologies, we refer the reader to dedicated books, such as~\cite{brand2015_book}.} 

\subsection{Energy Harvesting via non-RF Sources}
\label{sec:ext}

Energy Harvesting (EH) processes refer to the set of techniques used to capture electrical energy from ambient sources, thanks to dedicated conversion circuits, to be used to extend the lifetime of battery-powered wireless devices~\cite{brand2015_book},\cite{Moser2008}.

EH capabilities are appealing for a variety of applications, especially the ones involving pervasive devices, such as \acp{WBAN}, \acp{WSN}, and \ac{IoT}. Indeed, from the networking perspective, one of the main challenges of these applications domains has been always the limited energy availability provided by on-board batteries, as well as the high maintenance costs deriving by batteries replacement operations. On the one hand, EH capabilities can provide additional energy availability for battery-powered devices, enabling enhanced services (e.g., higher throughput, increased devices availability). On the other hand, EH circuits could replace batteries, possibly reducing operational costs, the form factor of the devices, and their impact on the surrounding context. 

Several environmental or non-dedicated sources can be used to scavenge energy and to power embedded circuits, enabling wireless networks \cite{chalasani2008}, \cite{sherazi2018_adhoc}. An overview of the main features of these sources is provided below, while a summary of the harvesting capabilities of the several sources is provided in Table~\ref{tab:energy}.
\begin{table}[htbp]
\color{black}
\centering
\caption{Overview of Energy Harvesting Sources and related Power Densities.}
\begin{tabular}{|P{2cm}|P{3.5cm}|P{1.7cm}|}
\hline
\textbf{EH Source Type} & \textbf{Power Density} &\textbf{Ref.} \\
\hline \hline
Solar & $\left[ 0.006 - 15 \right]$ $mW/cm^{2}$ & \cite{prauzek2018energy} \\
\hline
Wind & $\left[ 0.065 - 28.5 \right]$ $mW/cm^{2}$& \cite{sudevalayam2011, Habib2017} \\
\hline
Mechanical - Piezoelectric & $\left[ 0.11 - 7.31 \right]$~ $mW \cdot g^2/cm^{3}$ & \cite{brand2015_book} \\
\hline
Thermoelectric & $\left[ 15 - 60 \right]$ $\mu W/cm^{3}$ & \cite{prauzek2018energy}\\
\hline
RF & $\left[ 1.2\cdot 10^{-5} - 15 \right]$ $mW/cm^{2}$ & \cite{kim2014,prauzek2018energy}\\
\hline
\end{tabular}
\label{tab:energy}
\end{table}

\textbf{Solar Sources.} 
Historically, solar-to-DC conversion techniques have been the most investigated ones in the research community. The devices powered through solar energy use solar panels to convert light into energy. Due to the intermittent presence of the solar source (e.g., due to the adverse weather conditions or during darkness periods), rechargeable batteries and capacitors are used to store the excess energy and to guarantee continuous energy availability. We refer the interested readers to the valuable contribution in~\cite{Raghunathan2005_ipsn} for an overview of the most important challenges and hardware design considerations about solar-powered embedded systems. Generally speaking, the latest scientific sources available in the literature show that solar cells for embedded systems can achieve a power density in the range $\left[ 0.006 - 15 \right]$ \textcolor{black}{$mW/cm^{2}$} per time unit, depending on the luminosity level~\cite{prauzek2018energy}.

\textcolor{black}{
\textbf{Mechanical Sources.} Natural mechanical vibrations are among the most popular and attractive sources for energy harvesting. Indeed, mechanical vibrations generated from natural events (such as seismic motion and water tides) produce kinetic energy, characterized by a power density that is significant enough to power embedded devices~\cite{brand2015_book}. The systems leveraging mechanical sources for energy harvesting are usually classified based on the physical principle used to transduce from kinetic to electrical energy, and they can be partitioned into piezoelectric, electromagnetic, and electrostatic systems. Piezoelectric systems work according to the \emph{piezoelectric effect} first described by Marie Curie in 1880, by using the electric charge produced within a material as a consequence of a mechanical stress~\cite{alvarado2012energy}. These systems often use dedicated materials, such as the zinc oxide and polymer-based materials, that are well-known to be characterized by relevant electromagnetic coupling. Electromagnetic harvesters leverage the \emph{electromagnetic induction} principle, by generating voltage as a result of the variation of the magnetic field. Compared to piezoelectric-based systems, they are more difficult to integrate with modern Micro-Electrical Mechanical Systems (MEMS), but in medium-size products, they exhibit a suitable behavior, especially for low-frequency applications. Conversely, electrostatic harvesters are particularly suitable for IoT sensors. This latter class of harvesters leverages the \emph{variable capacitance} principle, generating different voltages by changing the capacitance of the system. This is usually achieved by oscillating a mass attached to a plate, thus varying the distance of this element from another plate. This class of harvesters is particularly suitable for embedded integration. However, these systems usually require an external voltage source, necessary to charge the plates, and they generate a limited amount of power. \\
Without loss of generality, mechanical EH systems generate power proportionally to the weight of the harvester (e.g., the mass in electrostatic harvesters)~\cite{pozo2019}. Therefore, when integrated with wireless devices, on the one hand the mass of the harvester should be maximized while, on the other hand, satisfying the specific constraints on the size of the system. Given that the production of the energy can be disjoint from its usage, the energy scavenged from mechanical sources is typically stored in rechargeable batteries, exhibiting slower discharge processes than capacitors. Several applications and even commercial products using a mechanical source for EH are available on the market~\cite{elvin2013}. The hardware details regarding mechanical energy harvesting can be found in~\cite{brand2015_book}. The same source indicates that piezoelectric EH systems generate a normalized power density in the range $\left[ 0.11 7.31 \right]$~ $mW \cdot g^2/cm^{3}$, while recent works such as~\cite{sherazi2018_adhoc} report values of $184$~$\mu W/cm^{3}$ at $10$~Hz for electrostatic harvesters and $0.21$~$\mu W/cm^{3}$ at $12$~Hz for electrostatic harvesters. These values have been also reported in the updated Table~\ref{tab:energy}.
}

\textbf{Wind Sources.} Similarly to the renewable energy sources previously discussed, wind sources for EH have been widely investigated in the literature. These systems use wind micro-turbines as the mechanical source enabling movement and, thus, AC generation.
\textcolor{black}{More in detail, the movement of the magnets on the stator of the micro-turbine generate AC electricity, that is then converted into DC using a bridge rectifier.}
However, compared to solar and additional mechanical EH sources, wind EH systems often require bulky equipment, very hard to be integrated with embedded devices. In addition, the wind power source is generally more unreliable than the other ones, as wind phenomena are more intermittent and time-varying. More details on hardware-related trade-offs and integration can be found in~\cite{Habib2017}. 
\textcolor{black}{The same source indicates that wind micro-turbines suitable for embedded systems can achieve a power density in the range $\left[ 0.065 - 28.5 \right] mW/cm^2$ per time unit, assuming a wind speed of $6.3 m/s$ and a turbine with a length of $30 cm$. Overall, we notice that the power density of wind EH sources depends strictly on (i) the wind intensity, (ii) the air density, (iii) the wind speed, and (iv) the length of the turbine rotor blade~\cite{Habib2017}.}

\textcolor{black}{\textbf{Thermoelectric Sources.} These sources have been historically used mainly in the first space and terrestrial applications of EH, and they are recently expanding also in other contexts, such as underwater~\cite{buckle2013}. Overall, systems using EH via thermo-electric sources leverage the temperature difference between two poles to generate energy. Recently, their usage has been proposed also for embedded devices, especially for \acp{WBAN} and micro-structured devices, thanks to the frequent temperature difference involved in such scenarios. Overall, thermoelectric power harvesters are considered more reliable and long-lasting than other EH sources, especially the mechanical ones, because of the relatively reduced exposition to movements and mechanical shocks. The hardware considerations regarding thermoelectric generators can be found in the contributions by the authors in~\cite{Alhawari2013}. The latest scientific sources available in the literature indicate that thermoelectric harvesters used with embedded systems can reach a power density level in the range \textcolor{black}{$\left[ 15 - 60 \right]$ $\mu W/cm^{3}$} per time unit, depending on the specific thermoelectric source~\cite{Habib2017}.}

We highlight that all the above-described sources leverage \emph{green} sources, i.e., natural phenomena not requiring any pre-existing system deployment. This is different from RF harvesting, where existing network infrastructure is adopted. Wireless charging methods are described in the following section.

\subsection{Energy Harvesting via RF Sources}
\label{sec:rf_charge}

The energy to power an embedded system can be obtained also via electromagnetic power transfer. This is the most promising and attractive technology for embedded devices, as the power generation process takes place via wireless links, without any physical connection between the energy source and the powered device. More in detail, embedded devices powered via electro-magnetic sources use dedicated circuits, able to extract power from the surrounding electromagnetic field, thanks to inductive, capacitive, and magneto-dynamic coupling principles. We refer the interested readers to the contributions by the authors in~\cite{bi2016wireless} for an overview of the hardware principles driving electromagnetic energy harvesting. The latest scientific sources available in the literature indicate that RF signals used to power embedded devices can achieve a power density in the range \textcolor{black}{$\left[ 1.2\cdot 10^{-5} - 15 \right]$ $mW/cm^{2}$} per time unit, depending on factors such as the distance from the RF source and the peak power of the signal~\cite{kim2014, prauzek2018energy}.

Overall, three main electromagnetic wireless powered network architectures can be identified: (i) \aclp{ABCN} (ABCN); (ii) \aclp{WPCN} (WPCN); and, (iii) \aclp{SWIPT} (SWIPT). The architectures and networking considerations crucial for the description of security issues and services connected to these technologies are described in sections~\ref{sec:back}, ~\ref{sec:wpcn}, and~\ref{sec:swipt}, respectively.

\subsection{\aclp{ABCN} (ABCN)}
\label{sec:back}

The devices resorting to ambient backscatter for power supply rely on \ac{RF} signals already available in the environment to gain energy. These RF sources include WiFi, Cellular, Radio, and TV Broadcast, i.e., signals that are used for tasks other than the specific power supply of the device. From the hardware perspective, dedicated antennas are used to convert such signals in a small portion of power (typically in the order of $\mu$Watts), used to reflect the signal with encoded data. In turn, other ABCN devices in the network use the same principle to reply and enable D2D communications~\cite{Vanhuynh2018}.

\begin{figure}[htbp]
    \centering
    \includegraphics[width=\columnwidth]{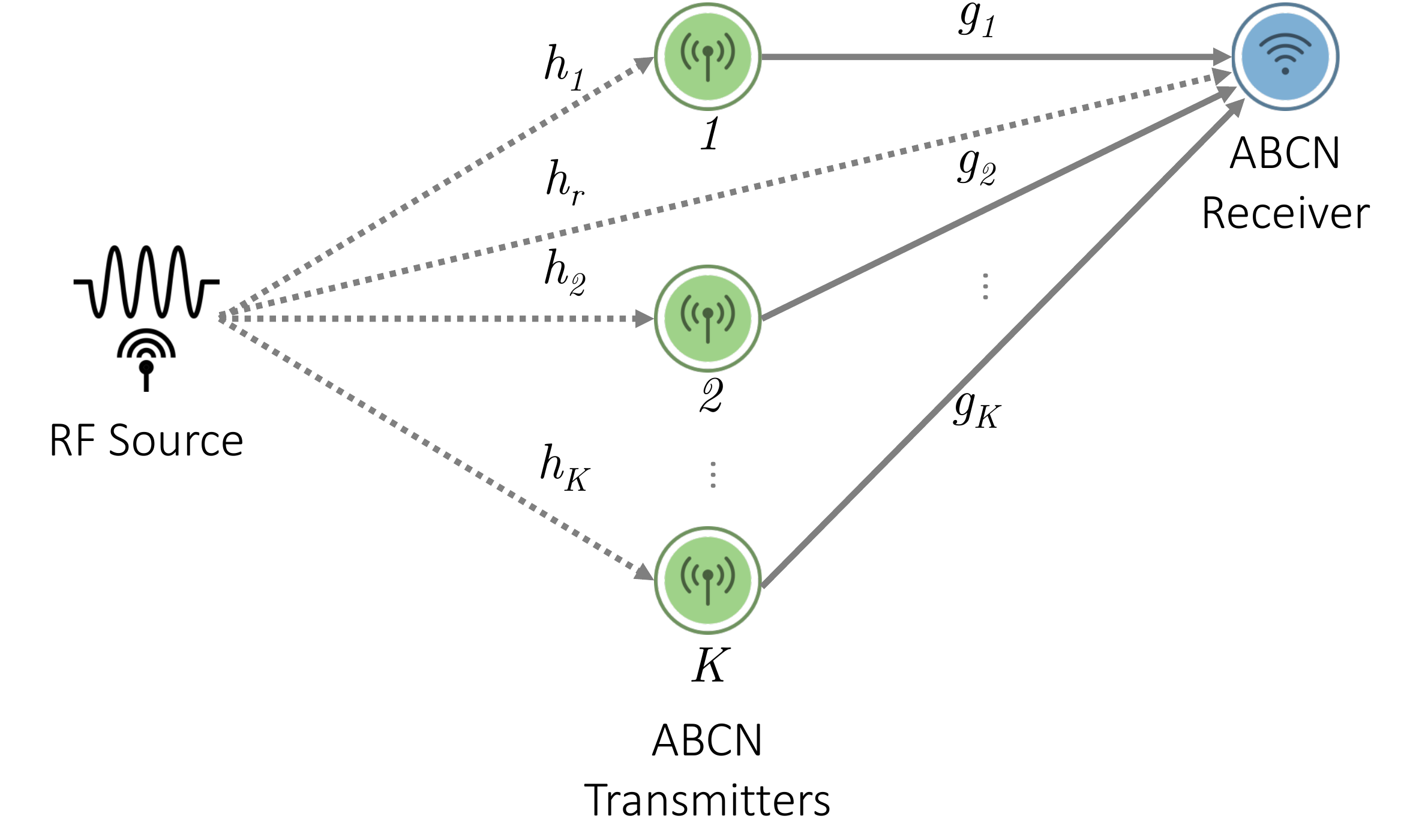}
    \caption{System Model of a generic \acl{ABCN}.}
    \label{fig:abcn}
\end{figure}

Let us assume a generic ABCN network, as the one represented in Figure~\ref{fig:abcn}, where a generic RF source is used as the ambient RF signal powering $K$ ABCN Transmitters to deliver information to a single ABCN Receiver.

Denoting with $h_k$ the channel between the RF source and the $k-th$ ABCN transmitter, and with $g_k$ the channel between the $k-th$ ABCN transmitter and the ABCN receiver, the signal $y(n)$ related to the $n^{th}$ bit of the message received by the ABCN receiver can be modeled as in the following Eq.~\ref{eq:abcn}, as the sum of the direct component from the RF source and the reflections from the $K$ ABCN transmitters.
\begin{equation}
    \label{eq:abcn}
    y(n) = h_r \cdot s (n) + \sum_{k=1}^K h_k \cdot g_k \cdot \eta_k \cdot s_k(n) + e(n),
\end{equation}

where $s_k(n)$ denotes the complex baseband signal originated by the $k-th$ ABCN transmitter, $e(n)$ is channel noise, modeled as \ac{AWGN}, and, finally, $\eta_k$ is the \emph{backscatter efficiency factor}, being $\eta \in \left[ 0,1 \right]$, indicating the backscatter efficiency of the $k-th$ transmitter. 

\subsection{\aclp{WPCN}}
\label{sec:wpcn}

\begin{figure}[htbp]
    \centering
    \includegraphics[width=\columnwidth]{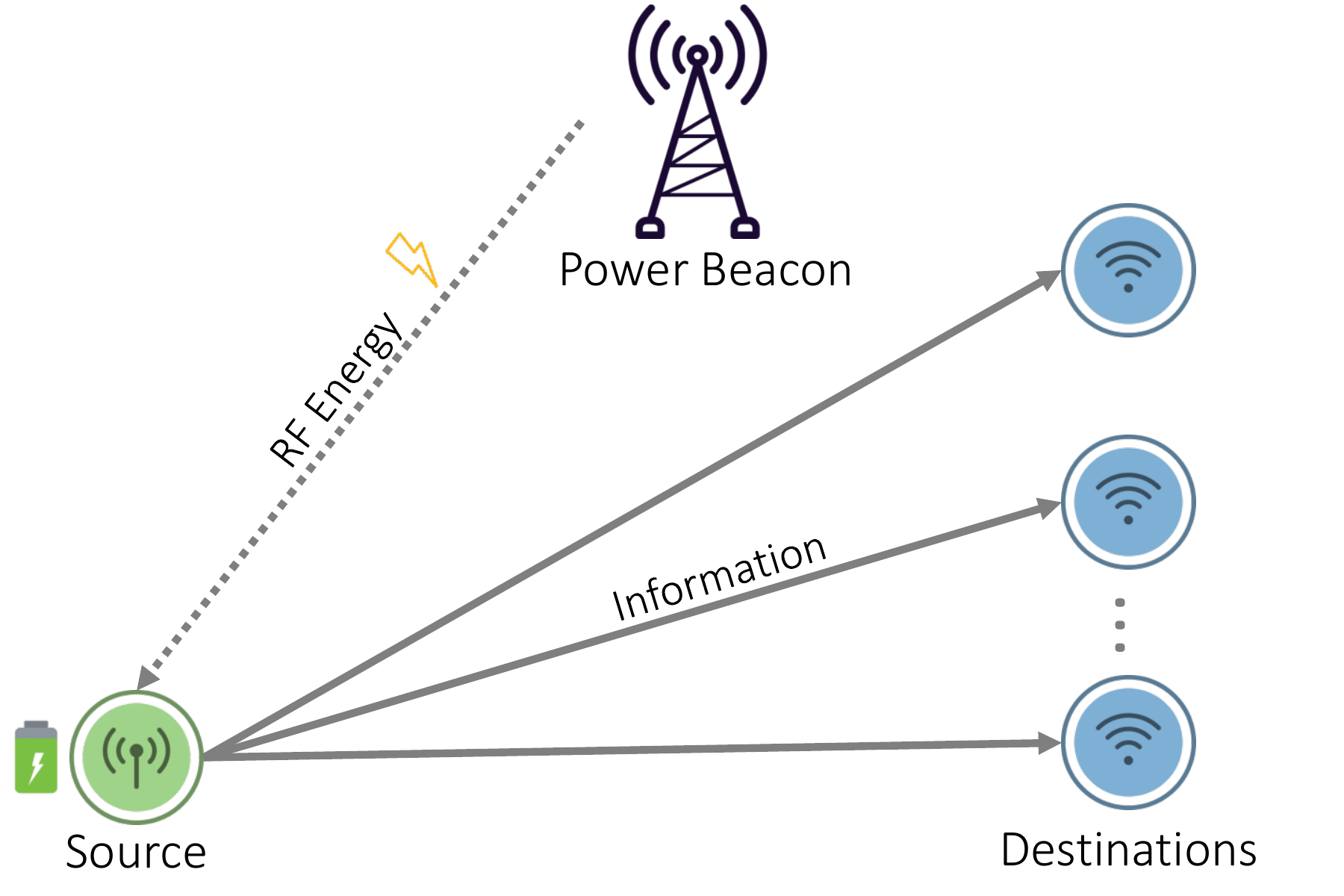}
    \caption{System Model of a generic \acl{WPCN}. The power beacon emits RF source that powers the transmitter and enable it to communicate with a receiver.}
    \label{fig:wpcn}
\end{figure}

\aclp{WPCN} (WPCN), often referred to as \emph{Far-Field \ac{WPT} networks}, are communication networks enabled through the presence of dedicated power sources, providing the necessary energy to power the devices via the \ac{WPT} technology. An exemplary system model is depicted in Figure~\ref{fig:wpcn}, where a power beacon transmits a dedicated RF signal that is used by the source to accumulate the necessary energy to communicate with multiple destination devices.

Compared to backscatter communication networks, WPCNs use dedicated signals (not used for other communication tasks), characterized by larger communication ranges, thanks to larger transmitting antennas compared to the radiation wavelength~\cite{perera2018_comst}. 

Compared to \ac{SWIPT} working according to the \ac{TS} architecture (see Section~\ref{sec:swipt}), instead, the main difference lies in the fact that the harvested energy is used to enable the transmission of information rather than the re-charge of the devices, thus anticipating the communication phase (this communication paradigm is often referred as \emph{harvest-then-transmit}). Moreover, the signal that powers the devices is decoupled from the information signal, as it is not specifically meant to transfer information.

Assuming that $\tau_0$ is the amount of time dedicated to energy harvesting in a block duration of time $T$, such energy harvesting time should be optimized to trade-off between the power available at the device and the data rate, being these metrics directly related to the \ac{SINR} at the destination and thus to the secrecy capacity of the communication link (see Sec.~\ref{sec:swipt}). Moreover, severe security issues at the physical layer arise especially for devices located far from the power source, that can harvest less energy than the others but require more power to push the transmission farther.

We refer the interested readers to the valuable contribution by the authors in~\cite{bi2015} for more details about the networking issues characterizing WPCNs, while security-related scientific contributions will be investigated in Sec.~\ref{sec:phy_secrecy}.

\subsection{\ac{SWIPT}}
\label{sec:swipt}

\begin{figure}[htbp]
    \centering
    \includegraphics[width=\columnwidth]{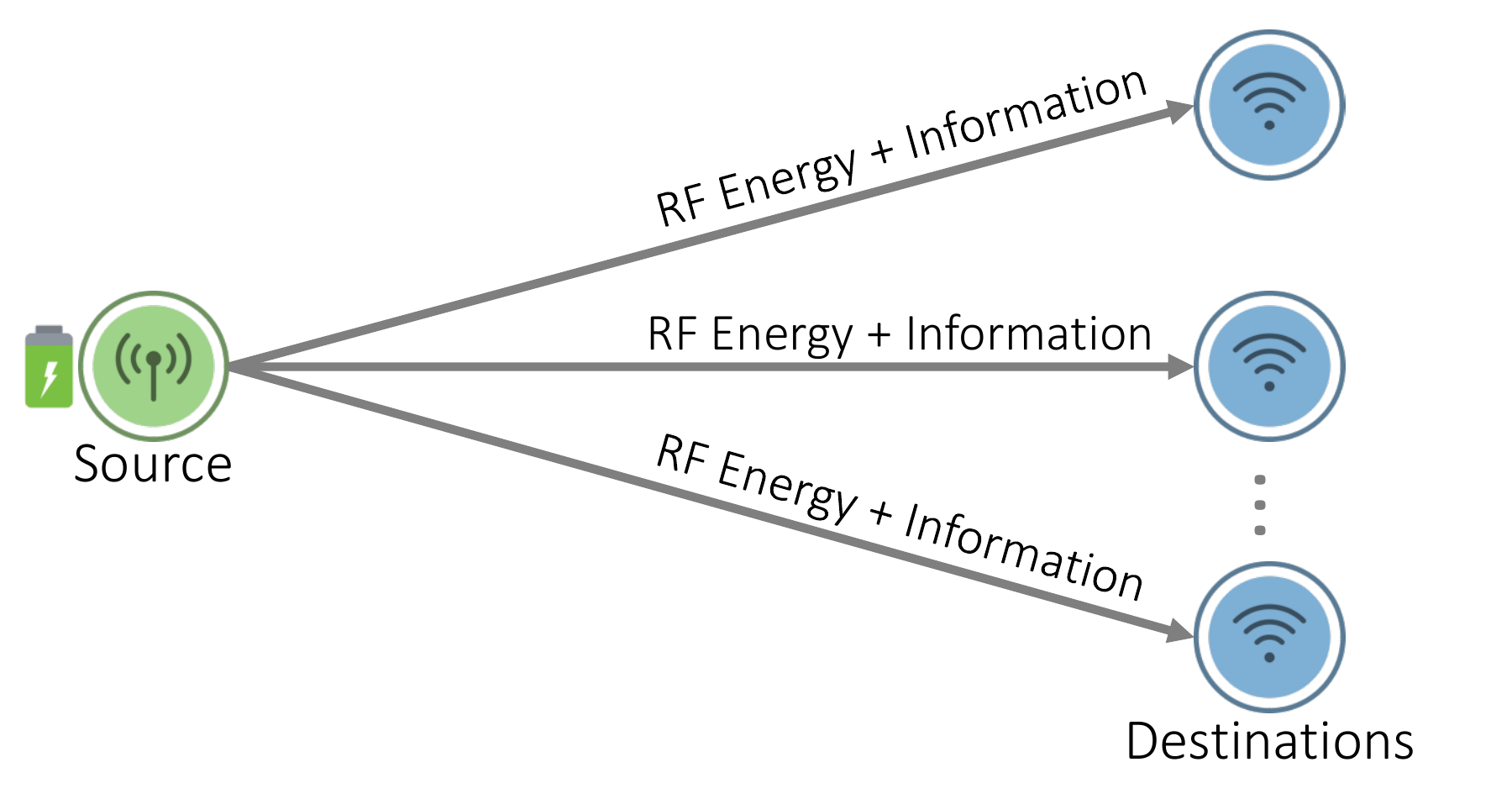}
    \caption{System Model of a generic \acl{SIMO} \acl{SWIPT} network. A battery-powered source emits a signal transferring information and RF energy, at the same time.}
    \label{fig:swipt}
\end{figure}

As shown in Figure~\ref{fig:swipt}, \acl{SWIPT} (SWIPT) networks
are characterized by contextual EH and information transfer processes, via dedicated hardware architectures. 

Generally speaking, \ac{SWIPT} systems are characterized by an emitting power station and several $N$ ($N >1$) of destination nodes. For the sake of simplicity, in the following, we introduce the notation for the case of a \ac{SIMO} system, where the multiple receiving entities are equipped each with a single antenna, while the extension for multiple-antenna systems is straightforward. 

The complex baseband signal transmitted by the source, namely $\mathbf{x}$, is given by the linear combination of the symbols emitted by the transmitting antennas to the receivers, as in the following Eq.~\ref{eq:tx}.
\begin{equation}
    \label{eq:tx}
    \mathbf{x} = \sum_{n=1}^N \mathbf{w_n} \cdot s_n,
\end{equation}

where $\mathbf{w_n}$ are the weights that are modulated by the transmitting antenna to the signal $s_n$ for the $n^{th}$ receiver, generally referred to as a \emph{beamforming vector}. 

The complex signal available at the generic $n^{th}$ receiving antenna, denoted as $y_n$, depends on both the transmitted signal $\mathbf{x}$ and the distortion introduced by the wireless channel, modeled by the matrix $\mathbf{h_n}$. Without loss of generality, the following Eq.~\ref{eq:rx} holds:
\begin{equation}
    \label{eq:rx}
    y_n = \mathbf{h_n}^H \cdot \mathbf{x} + e_n, 
\end{equation}

where the $H$ exponent denotes the Hermitian transformation of the matrix $\mathbf{h_n}$ (i.e., the transpose and conjugate of the matrix), while $e_n$ is a noise component at the $n^{th}$ receiving antenna, modeling undesired channel effects as \ac{AWGN} with variance $\sigma_n^2$. In addition, $\mathbf{h_n}$ is the distortion caused by the wireless channel, and it is frequently modeled according to a circularly symmetric complex normal distribution, having zero mean and noise covariance matrix, i.e., $\mathcal{S}$, namely $e \sim \mathcal{CN} (0,\mathcal{S}) $. In most of the works analyzed in this paper, the covariance matrix is statistically modeled according to the flat-fading Rayleigh model, while the Nakagami-$m$ model is used when aerial links are involved \cite{kermoal2002_jsac}.

The following manipulation of the signal at the receiver depends on the hardware architecture assumed for the receiving device. The literature mainly considers four hardware architectures: Time Switching (TS), Power Switching (PS), Separate Receivers (SR), and Antenna Switching (AS). The first three architectures are discussed in the following subsections, while we neglect the logic of the AS architecture given that none of the scientific contributions studied in this paper investigated this configuration (the interested readers can refer to the description in~\cite{Tran2017} for more details). 

\subsubsection{Time Switching (TS) Receiver Architecture}
\label{sec:ts}

Time-Switching (TS) EH receivers split the time duration of a time block $T$ in two distinct non-overlapping phases, lasting $\alpha_n T$ and $\left( 1 - \alpha_n \right) T$, where $\alpha_n$ is referred as the \emph{time-switching parameter}. In the first period, namely \emph{Information Decoding (ID) Phase}, lasting $\alpha T$, the receiver decodes information; then, in the remaining part of the time block, namely the \emph{Energy Harvesting Phase}, it uses the signal to harvest energy and charge the device. It is worth noting that, differently from the logic of a \ac{WPCN}, the energy harvesting phase follows the information decoding phase.

The signal available for ID at the receiver, namely $y^I$, can be expressed as in the following Eq.~\ref{eq:rx_ts_id}, where the signals emitted by the source toward other users are considered as interfering transmissions.
\begin{equation}
    \label{eq:rx_ts_id}
    \begin{split}
    y^I & = \sum_{n=1}^N \mathbf{h_n}^H \cdot \mathbf{x} = \\
    & = \mathbf{h_n}^H \cdot \mathbf{w_n} \cdot s_n + \mathbf{h_n}^H \cdot \sum_{j \ne n} \mathbf{w_j} \cdot s_j + e_n.
    \end{split}
\end{equation}

\textcolor{black}{
Thus, the \ac{SINR} for the ${n^{th}}$ user can be expressed as in the following Eq.~\ref{eq:ts_sinr}. 
\begin{equation}
    \label{eq:ts_sinr}
    SINR_n = \frac{| \mathbf{h_n}^H \cdot \mathbf{w_n} |^2}{\sum_{j \ne n} | \mathbf{h_n}^H \cdot \mathbf{w_j} |^2 + \sigma_n^2 }.
\end{equation}
}

The SINR can be used to obtain the capacity of the channel for each user, i.e., the maximum rate of information available at the user. According to the Shannon Theorem on channel capacity, it is defined as in the following Eq.~\ref{eq:ts_thr}.
\begin{equation}
    \label{eq:ts_thr}
    C_n = \alpha_n \log_2 (1 + SINR_n), n\in N.
\end{equation}

\subsubsection{Power Switching (PS) Receiver Architecture}
\label{sec:ps}

When the receiver works according to the Power Switching (PS) hardware architecture, the information decoding phase and the energy harvesting phase take place at the same time. This is achieved thanks to dedicated circuit design, splitting the received power signal in a part used for information decoding and another part used for EH, according to a \emph{power splitting factor}, namely $\rho_n$. As for the TS architecture, the information signal available at the receiver can be modeled as depicted in Eq.~\ref{eq:ps_id}.

\begin{equation}
\label{eq:ps_id}
\begin{split}
    y^I &= \sqrt{\rho_n} \left( \sum_{n=1}^N \mathbf{h_n}^H \cdot \mathbf{x} \right) + q_n = \\
     & = \sqrt{\rho_n} \left( \mathbf{h_n}^H \cdot \mathbf{w_n} \cdot s_n + \mathbf{h_n}^H \cdot \sum_{j \ne n} \mathbf{w_j} \cdot s_j + e_n \right) + q_n,
\end{split}
\end{equation}

\noindent
where $q_n$ models an additional AWGN due to the non-idealities introduced by RF-to-DC conversion circuits, having variance $\delta_n^2$. From the above expression, it is possible to obtain the SINR and the channel capacity for the generic user $n$, as in the following Eq.~\ref{eq:ps_sinr} and Eq. ~\ref{eq:ps_thr}, respectively.

\begin{equation}
    \label{eq:ps_sinr}
    SINR_n = \frac{ \rho_n | \mathbf{h_n}^H \cdot \mathbf{w_n} |^2}{ \rho_n \sum_{j \ne n} | \mathbf{h_j}^H \cdot \mathbf{w_j} |^2 + \rho_n \delta_n^2 + \sigma_n^2 }.
\end{equation}
\begin{equation}
    \label{eq:ps_thr}
    C_n = \log_2 (1 + SINR_n), n \in N.
\end{equation}

\subsubsection{Separated Receiver (SR) Architecture}
\label{sec:sr}

While TS and PS are different hardware architectures, the Separated Receivers (RC) is a scenario where some receivers are specifically dedicated to EH activities, while others are specifically dedicated to communication tasks. Defining $A$ the set of users dedicated to communication tasks and $B$ the ones dedicated to EH (with $|A|$ + $|B|$ = $N$), the signal transmitted by the source can be modeled as in the following Eq.~\ref{eq:sr_tx}:
\begin{equation}
    \label{eq:sr_tx}
    x = \sum_{n \in A} \mathbf{w_n} \cdot s_n + \sum_{m \in B} \mathbf{w_m} \cdot s_m,
\end{equation}
\noindent
where $s_n$ and $s_m$ are the information signal and the energy signal, respectively. 
Considering the generic $n^{th}$ user selected for information decoding, the signal received at this user is expressed by the following Eq.~\ref{eq:sr_rx}.

\begin{equation}
\label{eq:sr_rx}
\begin{split}
    y_n^I &= \sum_{n=1}^N \mathbf{h_n}^H \cdot \mathbf{x} = \\
     &= \sum_{n \in A} \mathbf{h_n}^H \mathbf{w_n} s_n + \sum_{m \in B} \mathbf{h_m}^H \mathbf{w_m}  s_m + e_n = \\
     &= \mathbf{h_n}^H \mathbf{w_n} s_n + \sum_{j \in A, j \ne n} \mathbf{h_j}^H \mathbf{w_j} s_j + \sum_{m \in B} \mathbf{h_m}^H \mathbf{w_m} s_m + e_n.
\end{split}
\end{equation}

As in the previous hardware designs, the information signal can be used to derive the SINR and the channel capacity for the scenario, as in the following Eq.~\ref{eq:sr_sinr} and Eq.~\ref{eq:sr_thr}, respectively.
\begin{equation}
    \label{eq:sr_sinr}
    SINR_n = \frac{ | \mathbf{h_n}^H \cdot \mathbf{w_n}|^2}{\sum_{j \in A, j \ne n} | \mathbf{h_j}^H \cdot \mathbf{w_j} |^2 + \sum_{m \in B} | \mathbf{h_m}^H \cdot \mathbf{w_m} |^2 + \sigma_n^2},
\end{equation}

\begin{equation}
    \label{eq:sr_thr}
    C_n = \log_2 ( 1 + SINR_n), n \in A.
\end{equation}

\subsection{Unique Features in EH Networks}
\label{sec:eh_chal}

Without loss of generality, we can notice that embedded devices powered via EH technologies exhibit unique energy availability features, severely influencing, in turn, the design of security solutions.

Overall, two main features can be highlighted. First, the availability of the specific source used to generate power supply is both intermittent and out-of-scale for embedded devices, as qualitatively shown in the following Figure~\ref{fig:supply}.
\begin{figure}[htbp]
    \centering
    \includegraphics[width=\columnwidth]{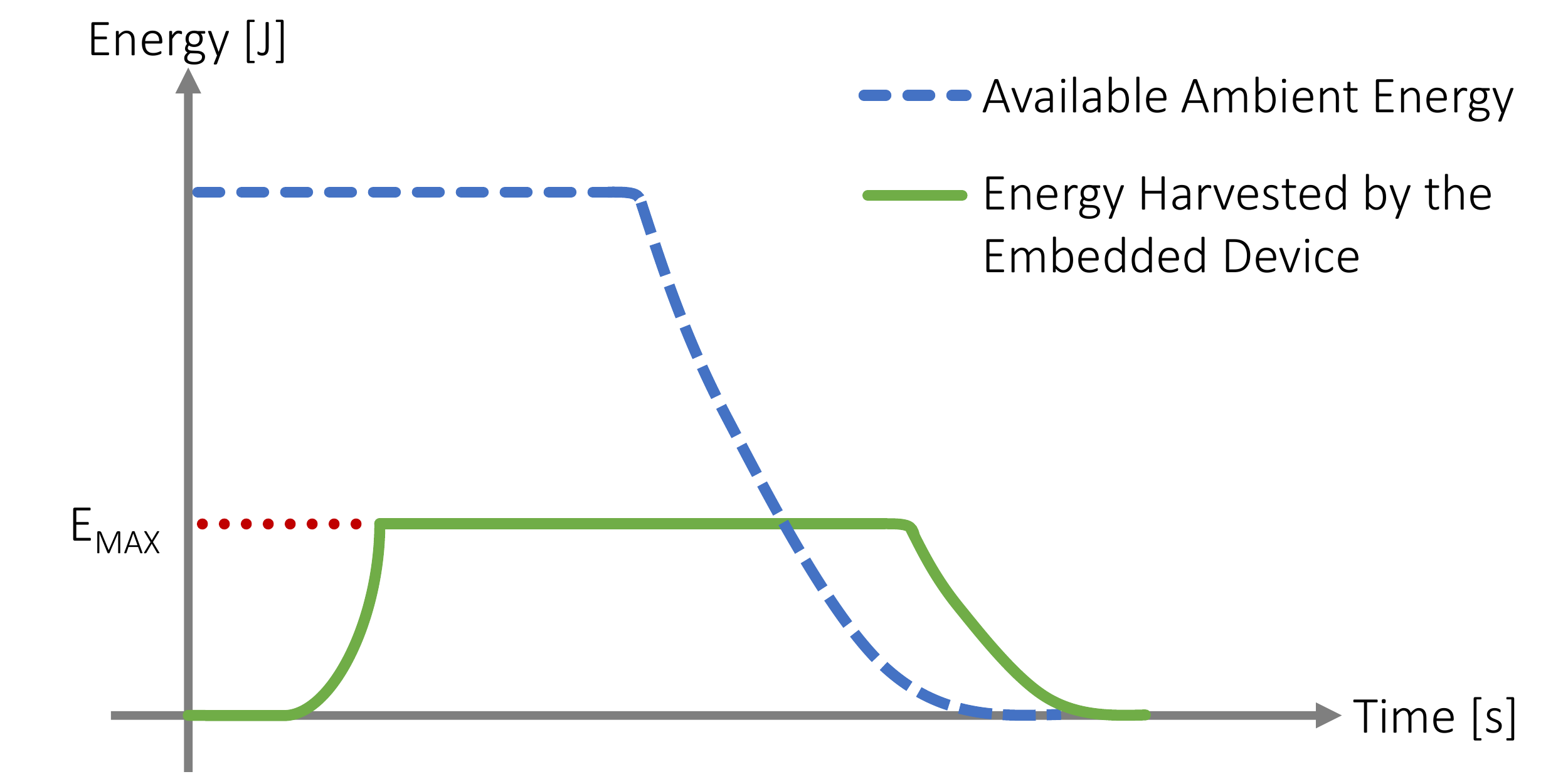}
    \caption{Qualitative description of the EH process. When high energy is available, the EH circuits power the capacitors on embedded devices, up to a maximum level of $E_{MAX}$. When the energy source is not available, the energy is quickly drained.}
    \label{fig:supply}
\end{figure}

The overall energy available for the re-charging is generally characterized by periods of high availability, alternated with periods of low/no availability. During the full availability periods, the EH circuits can harvest energy up to a maximum level, (indicated as $E_{MAX}$ in the Figure \ref{fig:supply}), while the remaining energy cannot be harvested. When the availability of the external source decreases, the capacitors or batteries on-board of the embedded system usually start losing energy after a little time, quickly exhausting the energy resource based on the current drain of the load system.

\textcolor{black}{In addition, the level of the energy harvested by the embedded system is usually strictly connected with the received level on the re-charging equipment. The higher the impact of the energy source on the EH circuit, the higher the energy recycled, and the higher the energy available for the system. Especially when considering electromagnetic EH, the received energy source level could have a severe impact on the overall harvested energy per time unit, and thus on its availability for the embedded device. These considerations have motivated several scientific contributions, such as~\cite{corona2018_wisee} and~\cite{jornet2012_icc}, that provided energy consumption models suitable for specific application scenarios involving EH, such as aircraft and \ac{WSN} applications, respectively.}

We want to remark that all these unique features are all taken into consideration when designing security solutions for EH networks.

\section{Attacks and Threats in Energy Harvesting Networks}
\label{sec:attacks}

The underlying mechanisms and strategies enabling EH principles expose wireless devices and networks to several attacks at different layers of the protocol stack. These threats extend beyond trivial eavesdropping attacks, and target the overall availability of the network, its reliability, and its dependability.

Different attacks have been identified and discussed in the literature, each requiring specific tools and features by the adversary and targeting specific enabling technologies. A summary of these attacks is provided in Table~\ref{tab:attacks_eh}, concerning the techniques and capabilities employed by the adversary.

\begin{table*}[htbp!]
\centering
\caption{Overview of Attacks affecting \acl{EH} Networks}
\begin{tabular}{|P{1.5cm}|P{2cm}|P{2.5cm}|P{3cm}|P{3cm}|P{3cm}|}
\hline
\textbf{Ref.} & \textbf{Attack Name} & \textbf{Target Enabling Technologies} & \textbf{Attack Class} & \textbf{Vulnerability} &  \textbf{Threat} \\ \hline\hline
\cite{liu2016_commag, perera2018_comst, vo2017} & Unauthorized Information Sniffing & WPCN, SWIPT & Eavesdropping & No encryption & Data leakage, passive information acquisition \\ \hline
\cite{Shakhov2013} & Flooding & WPCN & DoS & No native anti-malware and intrusion detection systems & Quick battery exhaustion \\ \hline
\cite{belmega2017_tifs}, \cite{belmega2017_icc}, \cite{rezgui2019}, \cite{hoang2015}, \cite{knorn2019_csl}, \cite{guo2017_twc}, \cite{luo2018_sensys}, \cite{niyato2015_icc} & Jamming & Generic EH network & DoS & No anti-jamming solutions & Throughput degradation, service interruption \\ \hline
%
%
\cite{liu2016_commag} & Beamforming Vector Poisoning & WPCN, SWIPT & DoS & No interference detection and anti-jamming solutions & Resources exhaustion \\ \hline
\cite{chang2017_wisec, chang2019_tons, pu2017} & Stealthy Collision & WPCN & DoS & No centralized coordination entities & Energy depletion and devices batteries exhaustion
\\ \hline
\cite{luo2018_sensys} & Deauthentication Deadlock & WPCN & DoS & No authentication and encryption & Connection interruption \\ \hline
\cite{nozaki2018_iccia, moukarzel2017} & Electromagnetic leaks, Power Consumption, Timing, Powering Leakages & SWIPT, WPCN & Side Channel & No tamper-resistant solutions & Sensitive data leakage \\ \hline
\cite{Yang2017_mobicom} & Back-scattered RF Reflection & ABCN & Side Channel & No physical shielding mechanisms & Sensitive data leakage \\ \hline
\cite{krishnan2019} &  Checkpoint Tampering, Power Interruption & Generic EH network & Device Tampering & No integrity, confidentiality, authenticity & Full control of power supply, infinite loops \\ \hline
\cite{luo2018_sensys} & Playback & WPCN & Replay & No freshness of cryptography materials & Malicious (re)transmissions of valid data \\ \hline
\cite{Luo2018} & Pilot Signal Spoofing & ABCN & Spoofing & No messages authentication & Unauthorized delivering of messages and energy \\ \hline
\cite{liu2016_commag} & Charging (Cheating, Leeching, Flooding) & WPCN & Spoofing & No authentication and throughput balance solutions & Energy depletion and battery exhaustion; malicious nodes collude to create packet collisions against legitimate nodes \\ \hline
\cite{luo2018_sensys}, \cite{liu2016_commag} & Impersonation & ABCN, WPCN, SWIPT & Man In The Middle & No authentication & Unrecognizable fake nodes \\ \hline
%
%
\cite{liu2016_commag} & Malware & WPCN & Malware infection & No suitable anti-malware solutions & Energy depletion, interruption of charging operation  \\ \hline
%
%
\end{tabular}
\label{tab:attacks_eh}
\end{table*}

\textbf{Eavesdropping.} First, as highlighted by several contributions such as~\cite{perera2018_comst}, \cite{liu2016_commag}, and \cite{vo2017}, RF EH networks are usually deployed without any built-in encryption mechanism, exposing the delivered energy and information to eavesdropping attacks. Indeed, given that RF EH networks are usually characterized by reduced energy availability compared to other EH sources, integrating off-the-shelf encryption techniques is often not suitable. These considerations have motivated a significant literature production, both considering physical-layer security schemes to improve data secrecy and suitable cryptography techniques, tailored to devices with reduced energy availability. These research areas are discussed in details in the following sections~\ref{sec:phy_secrecy} and ~\ref{sec:light_crypto} of this paper.

\textbf{DoS}. Moreover, the literature provides several descriptions and practical examples of \ac{DoS} attacks, aimed at disrupting the availability of EH networks. In this context, networks powered via RF signals are particularly exposed. 
In this context, one of the simplest strategies to quickly exhaust the energy provision of EH devices is to realize \emph{energy-depletion attacks}. As highlighted by the authors in~\cite{dimauro2012_peccs}, an \emph{energy depletion} attack is a form of \ac{DoS} attack, where the adversary sends bogus packets to a wireless-powered node. The radio and processing operations required to analyze and reject these bogus packets lead to a quick drain of the device's battery, preventing any further operation of the node in the network. An example of energy depletion is the flooding attack discussed by the authors in~\cite{Shakhov2013}. They considered several attack scenarios where, for instance, the adversary could: (i) introduce a malware in a \ac{WSN} to increase the transmission power (and thus, the energy consumption) of legitimate devices; (ii) adopt \emph{jamming} techniques to introduce noise on the communication channel, to require the nodes to increase the power transmission and increase the energy consumption; and (iii) exploit the weakness of some traffic aggregation scheme, to increase the size of the re-transmitted packet.

When the transmitting devices use beamforming techniques to increase physical-layer security, an attacker can launch \emph{Beamforming Vector Poisoning} attacks, injecting a malicious signal on the same frequency used for legitimate communications~\cite{liu2016_commag}. This translates into a jamming attack and produces destructive interference with the main communication link, reducing the amount of energy that the node can use either for energy charge or information decoding.

While being valid in the context of \acp{WSN}, the above considerations apply also to the IoT, as discussed by the authors in~\cite{luo2018_sensys}. In this reference contribution, the authors highlighted the feasibility of most common attack techniques, such as jamming, replay, and spoofing towards the IoT constrained devices. Moreover, they introduced the \emph{de-authentication deadlock} attack. It is a type of \ac{DoS} attack, affecting wireless nodes authenticated to an access point. In this scenario, a malicious user could spoof the victim device and inject unauthorized de-authentication frames directed to the access point, forcing the disconnection of the victim. Specific protection technologies to defeat de-authentication deadlock attacks are discussed in Section~\ref{sec:defense}.

Other attacks at the physical layer of a \ac{WPCN} are discussed by the authors in~\cite{liu2016_commag}. Specifically, despite jamming can be used also for dedicated network protection purposes (see Section~\ref{sec:jamming}), malicious adversaries can use this attack vector to poison the communication channel used for the re-charging of the devices, leading to a \ac{DoS}. Several contributions in the literature highlighted the issues related to the jamming attacks against a system with energy harvesting. With specific reference to the control theory, the authors in~\cite{knorn2019_csl} analyzed the effects of the remote state estimation in a closed-loop system under the jamming attacks, taking into account the communication channel properties. Further, they study how jamming affects the performance in a scenario where energy harvesting is employed for battery-powered sensors (transmitter-side), by deriving a policy for the optimal energy allocation. Similarly, the authors in \cite{niyato2015_icc} considered a \ac{WPCN} where an attacker intercepts the energy transmitted to a legitimate user. They provided a theoretical model integrating game theory concepts, and they leveraged this model to optimize the energy required for the data transmission, and the attack policy to achieve effective jamming attacks. 

\textbf{Side-channel}. Side-channel attacks allow an adversary to gain relevant information from the device, including secrets keys or private data, by leveraging physical-layer features such as electromagnetic and magnetic leaks, timing information, and energy consumption.
As discussed by the authors in~\cite{nozaki2018_iccia} and~\cite{moukarzel2017}, side-channel attacks are a threat also for \ac{EH} networks. For instance, correlating the instantaneous energy consumption and the electromagnetic leaks generated during cryptography operations can reveal the secret key computed by the node.

\textbf{Replay \& Spoofing.} As highlighted by some contributions such as~\cite{Yang2017_mobicom} and~\cite{Luo2018}, signals enabling \acp{ABCN} are not authenticated, and thus they can be easily replayed and spoofed by an attacker.
While the authors in~\cite{Luo2018} contextualized this attack in a \ac{WBAN}, a specific type of side-channel attack technique has been described also by the authors in~\cite{Yang2017_mobicom}, where a powerful software malware is deployed in the network. This malware extracts sensitive information from the victim, and it leverages a WiFi \ac{NIC} to force wireless devices to backscatter surrounding RF signals, thus enabling a side-channel used to stealthily transmit the information.

\textbf{Device Tampering.} Recently, the authors in~\cite{krishnan2019} introduced a side-channel attack specific to intermittent computing systems, such as EH devices. During a power loss period, these systems use to save the state of the program as checkpoint(s) into non-volatile memory. At the restoration time of the power supply, the checkpoints are used to recover the state of the system, to further proceed with the computations. Malicious adversaries can exploit the \emph{power interruption} as an attack vector, to: (i) read the checkpoint data; (ii) tamper the checkpoint after the system restore; and (iii) finally, execute a checkpoint replay attack, where the same checkpoint is restored infinitely by executing the same section of code.

\textbf{Malware Infection.} We recall that another way for the attacker to quickly drain devices' battery include typical malicious applications, deployed on purpose on the target device to send multiple energy requests, degrading the system throughput. An example has been provided by the authors in~\cite{Yang2017_mobicom} (previously described), where malware is used to force a specific behavior on the target victim. 

Other attacks specific to RF EH networks are the ones targeting the trust of the messages used to provide the RF energy to charge the devices. As discussed in Section~\ref{sec:techno}, WPCNs require specific \emph{pilot} signals emitted by dedicated power beacons, that physically enable the harvested device. However, these signals usually do not integrate any security service, such as authentication, integrity, and confidentiality. Thus, as summarized by the authors in~\cite{kang2015_commag}, several attacks can be launched. 

Looking at the transmission side, the pilot signal used for energy charging can be spoofed, leading to a \emph{Pilot Signal Spoofing} attack. This could happen also when the EH device specifically requests power, sending a beacon request packet. In such a scenario, the adversary can intercept this message and impersonate the power beacon, reducing the throughput and the lifetime of the EH device~\cite{Luo2018}.

\acp{WPCN} are also vulnerable to \emph{charging attacks}, that can be categorized in \emph{leeching}, (ii) \emph{greedy}, and (iii) \emph{cheating} charging attacks. Leeching attacks are physical-layer attacks where devices keep re-charging without sending explicit requests, i.e., by taking advantage of energy leakages when powering neighboring devices. This can affect the awareness of a power beacon, especially if this is interested in maintaining an accurate map of the energy availability of the EH nodes~\cite{liu2016_commag}.

Greedy charging attacks involve \emph{greedy devices}, continuously requesting energy from the source, and thus preventing other devices to request energy. 

Cheating charging attacks, instead, involve a compromised EH node, reporting inaccurate energy measurements to the power source, forcing it to adjust (increasing or decreasing) the reference power level. This can translate either in a \ac{DoS} or in an \emph{unauthorized energy charging} attack. In the former, the reduction of the power level of the RF source decreases the size of the area where the signal is received, possibly excluding some devices from receiving enough energy. In the latter, increasing the power level of the pilot signal increases the area where such signal is received, possibly enabling malicious EH devices, controlled by the adversary.

We also highlight that, as discussed by the authors in~ \cite{liu2016_commag}, the harvested energy is subject to sensitive changes due to the presence of people in the deployment area. This property could enable the attacker to infer sensitive information about the status of the area surrounding the physical environment where the EH network is deployed, such as the number of people in the area, their movement, and distribution over time, to cite a few, leading to severe privacy attacks.

\textbf{Cooperative EH Networks.} Finally, besides the above-described attacks, it is worth mentioning some specific attacks targeting the novel concept of \emph{Cooperative Energy Harvesting Networks}, highlighted by the authors in~\cite{kang2015_commag}.

In cooperative \ac{EH} networks, a generic source node $S$ transfers energy towards a destination node $D$, by leveraging one or more relay nodes $R$. These networks do not support any inherent security property hence, besides sharing the same vulnerabilities of traditional WPCNs, these networks are exposed to further threats. In particular, since energy messages do not satisfy implicitly the integrity property, these networks are exposed to \emph{energy state forgery} attacks, where an adversary can tamper the information about its energy state, and send the tampered information towards a cooperator for illicit goals. For instance, when a legitimate node requests energy to the attacker, this latter one can refuse to provide it, claiming an insufficient energy level.
Moreover, a compromised node can trigger \emph{repudiation of energy} attacks, denying to own a given amount of energy after the transfer process initiated by a cooperator. In this case, the compromised node acts as a \emph{black hole}, accumulating energy locally and reducing the energy availability on other cooperating devices. These attacks can be anticipated by \emph{energy cheating} attacks, where the compromised node obtains sufficient trust from the victim. First, the compromised node sends messages and operates in the network as a legitimate device, gaining the necessary trust from the network. Then, when the trust is established, it requests energy from other devices and exhausts their energy availability. Finally, cooperative EH networks could be also exposed to \emph{collusion attacks}, where multiple devices collude to deny the reception of energy from a target cooperator device, forcing it to repeat the energy delivering operations, quickly exhausting the available energy.

The techniques to mitigate the above attacks leverage either lightweight tailored cryptography solutions or physical-layer features. The details are provided in sections~\ref{sec:light_crypto} and ~\ref{sec:defense}, respectively.

\color{black}
\section{Cryptography Techniques for EH Devices and Networks}
\label{sec:light_crypto}

A significant branch of the works on security issues for EH networks focus on the application of cryptography techniques to EH devices, adapting the operations and the implementation of the security operations to the new wave of systems. Specifically, the scientific contributions in this area strive to optimize the implementation and the software/hardware architecture of well-known cryptography approaches, based on the features and requirements typical of energy-harvesting networks. The final aim is to come up with customized versions of cryptography techniques, characterized by either intermittent or less-demanding energy consumption, thus being suitable for integration in devices powered via harvested sources.

Compared to the large amount of scientific contributions investigating data secrecy issues at the physical-layer (see Section~\ref{sec:phy_secrecy}, cryptography approaches can guarantee the specific security service in a deterministic fashion, while physical-layer data approaches are characterized by a probabilistic security protection. At the same time, these approaches usually involve higher demand on the system, thus requiring the implementation of dedicated optimization mechanisms.

An overview of the most important systemic features of these proposals is provided in Table~\ref{tab:eh_crypto}.
Indeed, despite each of these proposals is specifically meant for a particular EH technique, we argue that the logic and strategies of these proposals can be fully re-used even for additional EH sources and networks, being a valuable alternative to physical-layer security schemes and leading to feasible cryptography algorithms for generic EH networks.
\begin{table*}[htbp]
\color{black}
\centering
\caption{Comparison between approaches contextualizing cryptography solutions in EH networks. \newline The symbol \emph{N/A} indicates that the specific feature is not applicable.
}
\begin{tabular}{|P{0.8cm}|P{1.4cm}|P{2.4cm}|P{1.5cm}|P{1.7cm}|P{1.9cm}|P{1.5cm}|P{3.2cm}|}
\hline
\textbf{Scheme} & \textbf{Scenario} &\textbf{Target Security Service} & \textbf{Solution Type} &  \textbf{Assessment Methodology}  & \textbf{Hardware Board} & \textbf{EH Source} & \textbf{Proposed Technique} \\
\hline \hline
\cite{taddeo2010_secrypt}, \cite{mao2019_commlett} & IEEE 802.15.4 & Authentication, Confidentiality, Integrity & Software & Simulations & N/A & Generic & Security service dynamic adaptation \\ 
\hline
\cite{copello2018} & IMD - WSN & Confidentiality & Software & Real-Devices Experiments & WISP & RF Signal & Security service dynamic adaptation \\ 
\hline
\cite{Fang_2018_mdpi} & IoT & Confidentiality & Software & Simulations & N/A & RF signals & Security service dynamic adaptation \\ 
\hline
\cite{schaumont2016} & WSN & Authentication & Mixed Hardware - Software & Real-Devices Experiments & MSP430 & Solar cells & Security service dynamic adaptation \\
\hline
\cite{pellissier2011} & WSN & Confidentiality & Mixed Hardware - Software & Emulation & MSP430 and the AVR ATmega128l & Generic & Pre-computation techniques \\ %
\hline
\cite{bianchi2013_adhoc} & IEEE 802.15.4 & Confidentiality, Access Control & Software & Real-Devices Experiments & TelosB & Photovoltaic Cells & Pre-computation techniques \\ 
\hline
\cite{Suslowicz2017} & IEEE 802.15.4 & AES & Mixed Hardware - Software & Emulation & MSP430, ARM-Cortex M4 & Generic & Pre-computation techniques \\ %
\hline
\cite{lv2018_access} & WSN & Digital Signature & Software & Real-Devices Experiments  & MSP430 & Solar cells & Pre-computation techniques \\ 
\hline
\cite{Ateniese2017_tecs} & IEEE 802.15.4 & Generic Exponentiation Operations & Software & Real-Devices Experiments & MagoNode++, TelosB, and MICA2 motes & Photovoltaic cells and micro wind turbines & Pre-computation techniques\\
\hline
\cite{ateniese2017_vtc}& IEEE 802.15.4 & Generic Exponentiation Operations & Software & Emulation & TelosB & Generic & Offloading to energy-supplied devices \\ 
\hline
\cite{ellouze2015_sac} & WSN & Mutual Authentication & Software & Simulations  & N/A & Generic & Implementation optimizations for reduced energy consumption \\
\hline
\cite{Ellouze2013} & WBAN & Mutual Authentication & Software & Simulations  & N/A & Electro-cardiac signals & Implementation optimizations for reduced energy consumption \\
\hline
\cite{nozaki2018_iccia} & WSN & Encryption & Mixed Hardware - Software & Real-Devices Experiments & MonoStick &  Solar Cells & Implementation optimizations for reduced energy consumption\\
\hline
\cite{varga2015_iotj} & IEEE 802.15.4 & Authentication and Encryption & Software & Real-Devices Experiments & GreenNET & Photovoltaic cells & RF Operations Optimization \\ 
\hline
\cite{dimauro2015_precedia} & WSN & Key Agreement & Software & Simulations & N/A & Generic & Optimization based on the energy available on neighboring devices \\
\hline
\cite{valea2019} & IoT & Secure Context Saving & Hardware & Emulation & Not Available & Generic & Dedicated cryptography techniques for context saving \\  
\hline
\cite{Hylamia2018_wisec} & WSN & Hashing, Encryption & Hardware & Real-Devices Experiments & EHS Platform & Solar Panel & Implementation optimizations for reduced energy consumption\\ 
\hline
\cite{mohd2018_access} & IoT & Block Ciphers & Software & Simulations & N/A & Generic & Implementation optimizations for reduced energy consumption\\
\hline
\end{tabular}
\\
\label{tab:eh_crypto}
\end{table*}

Adapting dynamically the provied security service is one of the earliest strategies realized to cope with the limited energy availability of EH devices, and it has been conceived by the authors in~\cite{taddeo2010_secrypt}, in the context of IEEE 802.15.4 networks. They provided a software solution enabling a constrained WSN device to change at run-time its communication security settings, adapting dynamically to the charge level and increasing the lifetime. Besides, they designed their system to reduce the number of delivered packets (and thus, the energy consumption) when scarce energy is available. They contextualized their proposal and evaluated the energy and lifetime gain via simulations. Recently, the authors in~\cite{mao2019_commlett} tackled a similar scenario, contextualizing the problem in the novel \ac{SDN} computing paradigm.

\noindent
The authors in~\cite{copello2018} proposed a lightweight data scheduling strategy, trading off between data security and energy efficiency. Specifically, the proposed scheduling algorithm includes two security primitives, i.e., a stronger but energy-hungry one and a weaker but energy-efficient one. Depending on the available energy on the wireless platform, for each batch of data, the scheduler selects the most suitable, minimizing the impact of secure data transmission on the lifetime of the EH platform.

\noindent
The authors in~\cite{Fang_2018_mdpi} studied energy consumption optimization strategies to balance between security processing and energy harvesting. Considering that during RF harvesting periods the EH device cannot transmit or receive, and that the availability of the RF source is intermittent, the authors compared the effectiveness of either transmitting more small packets or less long packets, as well as running security algorithms, considering the harvesting availability of the devices. 

\noindent
A valuable contribution towards the identification and definition of the energy requirements for the authentication protocols in EH devices has been provided by the authors in~\cite{schaumont2016}. The authors evaluated the energy consumption of $18$ different authentication algorithms, measuring the energy expenditure with different software configuration and hardware optimizations. Using a solar-powered MSP430 hardware platform, they confirmed via experimental tests that energy efficiency can be improved by including software optimizations, as well as by a dynamic adaptation of the security level. A similar study has been also provided recently by the authors in~\cite{vracar2019_hindawi} on a PIC micro-controller.

Many contributions, such as \cite{pellissier2011} \nocite{bianchi2013_adhoc, Suslowicz2017, lv2018_access}--\cite{Ateniese2017_tecs} exploited pre-computation techniques. Indeed, as discussed in~\cite{cammarano2012}, EH networks are characterized by intermittent connectivity and charging periods, where time frames characterized by high energy availability are alternated with periods of scarce charging. In addition, EH circuits in WPCN devices always lose energy after reaching a given re-charge threshold, not fully exploring the advantages of EH capabilities. Thus, pre-computation techniques can be triggered during periods with high energy availability, to both use all the available energy and anticipate energy-demanding tasks, whose execution and energy cost could be too demanding in other time frames. The authors in~\cite{pellissier2011} investigated the adaptability of well-known cryptography algorithms to EH networks. Specifically, they worked on the algorithm part, and they found that pre-computing and storing a few key streams or parts of them can be useful when the system experiences a low charging level. Thanks to hybrid HW/SW optimizations, they optimized the CPU cycles to reduce the time and the energy required by cryptography operations, and they tested their methodology on real devices, such as the MSP430 and the ATMega128L micro-controllers.
Pre-computation techniques were also exploited by the authors in~\cite{bianchi2013_adhoc}, to allow devices equipped with both IEEE 802.15.4 communication capabilities and EH principles to integrate dedicated cryptography techniques, such as the \ac{CP-ABE} technique. In the proposed framework, namely AGREE, they provided specific optimizations to the CP-ABE cryptography algorithm, they pre-compute scalar multiplications during high-charge periods, and they use the CP-ABE technique to encode a session key instead of the data itself, translating an access control problem into an encryption one. They implemented their strategy in real constrained devices (the TelosB and MicaZ devices charged by photo-voltaic cells), showing the effectiveness and the reduced energy consumption of their methodology for EH networks. Similarly, pre-computation techniques have been used by the authors in~\cite{lv2018_access} to reduce the energy cost of digital signature algorithms by $44$\%, making them feasible for execution on a solar-cell powered micro-controller.

\noindent
Considering the heterogeneity in the harvesting capabilities of an RF EH network, the authors in~\cite{ateniese2017_vtc} designed HELIOS, a distributed protocol that allows network elements experiencing low energy availability to outsource the most energy-demanding operations to other nodes in the networks, being them power-supplied or with excess energy available. Considering also the presence of untrusted devices in the network, the authors discussed also specific strategies to reduce the impact of untrusted network elements, thanks to batch verification schemes. The authors extended their study in~\cite{Ateniese2017_tecs}, where they focused on the well-known \ac{ECDSA} protocol, used for authentication and integrity verification. Specifically, they coupled well-known ECC pre-computation techniques with the energy harvesting concepts previously introduced, optimizing the usage of excess energy and reducing energy waste due to the limited harvesting capabilities of embedded devices. Similar pre-computation optimizations were also provided by the authors in~\cite{Suslowicz2017}, where the authors referred to the excess amount of data generated during excess energy peaks as \emph{coupons}. Such coupons have been demonstrated to be a useful tool to speed up the execution of the \ac{AES} algorithm in CTR mode, as well as to speed-up hardware-based random numbers generation.

Other contributions, instead, addressed the energy shortage of EH networks by reducing the overall energy consumption of the security protocol. For instance, the authors in~\cite{ellouze2015_sac} designed a mutual authentication protocol for EH WSN that can split computationally-demanding operations among multiple periods, leveraging energy-aware checkpoints where the state of the system is saved and restored. Similarly, a protocol using electro-cardiac signals for harvesting and key generation has been proposed by the same authors in~\cite{Ellouze2013}.

\noindent
In~\cite{varga2015_iotj}, the authors address security issues as well. In particular, the authors introduced and described the novel GREENnet harvested hardware platform. This platform integrates application-layer security with the OSCAR architecture, which enjoys reduced energy consumption by decreasing the unnecessary handshake operations required by the standardized DTLS protocol.

\noindent
The authors in \cite{dimauro2015_precedia} focused on the key agreement security problem, and provided a strategy that allows optimizing the pairwise keys selection and distribution in an EH WSN. The optimization consists of balancing the number of reinforcement links and the availability of reinforcement neighbors based on the harvested energy available on the devices willing to share a key. 
Moreover, it is worth highlighting the recent disclosure of a hardware platform specifically dedicated to the testing of cryptography algorithms in EH-powered devices, recently presented by the authors in~\cite{Hylamia2018_wisec}. In their demonstration, they showed that a maximum charging time of $60$~s is necessary to execute the most energy-demanding cryptography scheme, i.e., HMAC-SHA. While this latter work provides action on the hardware perspective, the authors in~\cite{mohd2018_access} focused on the software part, optimizing the size and the operations required by a block cipher to reduce its energy consumption, making it feasible for EH devices.

\noindent
Finally, in the context of cooperative EH networks, the authors in~\cite{kang2015_commag} discussed some countermeasures against the attacks described in Section~\ref{sec:attacks}. They evaluated the inclusion of a registration authority and a trusted module in each device, to include additional security features in EH Cooperative Networks. High-level countermeasures such as the inclusion of period energy state reports via trusted modules, energy dual signatures via PKI-based methods, and the integration of symmetric and asymmetric encryption techniques are discussed. However, there is no evaluation of their impact on networks with a very limited budget, neither proposals to make their integration energy-friendly.

\subsection{Lessons Learned}
\label{sec:crypto_lessons}

\textcolor{black}{From the literature review summarized in the previous subsection, we notice that devices powered via non-RF sources (solar, wind, mechanical, thermoelectric) are usually characterized by higher energy availability per unit of time, if compared to systems powered via RF sources. Therefore, they can likely afford the execution of a limited set of cryptography primitives, optimized in such a way to meet the constraints of these devices.
\\
Without loss of generality, as the most important lesson learned from our study, three main research methodologies have been identified by the literature overview provided in the previous section.}

\textbf{Dynamic Security Service Adaptation.} Given that energy availability in EH devices is intermittent, one of the strategies highlighted in the literature is to provide a framework that acts in strict combination with an energy indicator. As soon as more energy becomes available, authentication and confidentiality are strengthened, while they are relaxed when the available energy reduces. Despite such schemes are valuable in balancing energy and security, they are particularly subject to jamming attacks. Indeed, it is enough for the adversary to reduce the charging profile of the device to lower the corresponding security level.

\textbf{Security Operations Optimization.} Following a similar logic to the one already used in constrained environments such as WSNs, other contributions cope with the energy limitation in EH networks by reducing the overall energy consumption of the cryptography protocols. This is achieved both by outsourcing the most demanding security operations and by optimizing the energy cost of some operations within block ciphers, often at the cost of a reduced security level. Despite being less vulnerable to jamming attacks, these strategies are often characterized by a reduced security level, not being suitable for deployment scenarios where adversaries can enjoy unlimited capabilities.

\textbf{Pre-Computation to Optimize Energy Usage in EH Networks.} One of the most effective strategies to provide security in EH networks is pre-computation. Indeed, pre-computation techniques take advantage of the periods when the device is fully powered, and anticipate the execution of the most energy-demanding tasks in these periods. Despite requiring significant engineering efforts on the cryptography protocols logic, pre-computation techniques allow maintaining the same security level of well-known solutions, successfully defeating both temporary jamming attacks and powerful adversaries. However, there are two main drawbacks. First, when the device experiences a prolonged period of reduced energy availability, the pre-computed values could be exhausted, leaving the device with no more energy to execute security tasks. Second, the design of pre-computation strategies should be validated by the scientific community, avoiding the creation of vulnerabilities in widely-accepted cryptography techniques.

\subsection{Future Directions}
\label{sec:crypto_future}

A few appealing research directions for cryptography solutions in the context of EH networks are provided below.

\textbf{Optimization Strategies Integration.} Despite the three strategies discussed in the previous subsections have received significant attention from the latest literature, still the integration between such strategies has not been explored. For instance, coupling the optimization of encryption operations with the dynamic adaptation of the security service could increase the minimum security level of the device. In this context, efficient frameworks are required to synchronize energy availability, expenditure, and security level tuning operations.

\textbf{EH Platforms Security Evaluations.} At the time of this writing, there is no scientific contribution experimentally validating the security of cryptography solutions on EH devices. Indeed, benchmarking the security of solutions tailored for EH devices, such as pre-computation techniques, is crucial to validate their strength and refine their security against evolving adversaries. In parallel, automatic security validation tools could help in establishing the security of either code optimization techniques and pre-computation strategies, avoiding the deployment of possibly weak security techniques.
\color{black}

\section{Data Secrecy Issues in Energy Harvesting Networks}
\label{sec:phy_secrecy}

Building on the very limited energy availability of EH networks and the general unsuitability of traditional cryptography techniques, several scientific contributions provided strategies to enforce physical-layer data secrecy in EH networks. \textcolor{black}{More in detail, the schemes discussed and cross-compared in Section~\ref{sec:phy_secrecy} implicitly assume that EH-powered devices are not able to run cryptography primitives, and thus they can neither encrypt data nor manage energy-consuming TCP-s connections. Thus, without any further protection, the related RF communications would be easily eavesdropped by a passive adversary. This is the reason motivating the intensive literature production intended to achieve security via non-cryptography schemes.}

These contributions largely focused on RF EH, for two main reasons. First, as highlighted by the comparison in Table~\ref{tab:energy}, RF EH harvesting techniques usually provide the least amount of power per time unit, justifying the assumption about the impracticality of cryptography techniques~\cite{Wang2019_ieeest}. Second, the research community has been attracted by the challenging trade-offs between energy, security, throughput, and reliability, typical of wireless-powered networks.

In this section, we first introduce the security objectives most used when investigating data secrecy at the physical-layer (see Section~\ref{sec:sec_obj}). Then, we describe the logic driving the introduction of relaying and friendly jamming strategies in RF EH networks (see Section~\ref{sec:relay} and Section~\ref{sec:jamming}). 
Without loss of generality, we notice that, despite investigating all the same security metrics discussed in Section~\ref{sec:sec_obj}, the contributions available in the literature are characterized by many differences, related to the assumed scenario, the adversarial model, and the specific objective.

To provide a meaningful and insightful overview of these works, we split the overall literature into three main categories, according to the specific adversarial model. Three adversarial models are mainly assumed: (i) external (Section~\ref{sec:ext_adv}); (ii) internal (Section~\ref{sec:int_adv}); and, (iii) untrusted relays (Section~\ref{sec:relay_adv}). 
The most important take-home messages arising from our discussion are summarized in Section~\ref{sec:secrecy_lessons}, while Section~\ref{sec:secrecy_future} highlights appealing future research directions in the area.

We remark that this section investigates and classifies only scientific contributions specifically tackling data secrecy security issues in EH networks. Thus, schemes focusing on EH network performance, not considering security issues, are out of scope for our work (indeed, exhaustive description and classification of these approaches can be found in dedicated surveys, such as~\cite{jameel2019} and~\cite{hossain2019_access}).

\subsection{Security Objectives}
\label{sec:sec_obj}

The operations of ABCN, WPCN, and \ac{SWIPT} networks can be optimized based on several design objectives, by selecting suitable beamforming vectors, or optimizing the transmission strategies and the deployment of the entities in the system. 

Several features are optimized in the literature, including the total transmission power at the transmitter, the energy expenditure at the receivers, the harvested power, and the throughput of the communication link.

In this manuscript, being focused on security issues, we are particularly interested in two security-related metrics, which are the \emph{secrecy rate} and the \emph{secrecy outage probability}.

Considering an eavesdropper $E$ is deployed in the scenario and it is interested in the correct decoding of information directed toward one of the legitimate devices, the \emph{secrecy rate} of the legitimate source-to-destination communication link at the physical layer is defined as in the following Eq.~\ref{eq:sec_rate}:
\begin{equation}
    \label{eq:sec_rate}
    S_n = C_n - C_e,
\end{equation}

where $C_n$ is the capacity of the legitimate channel and $C_e$ is the capacity of the channel source-eavesdropper~\cite{Aliberti2016}. 
A large part of contributions in the literature formulate proposals to maximize the secrecy rate (\ac{SRM} problem), by suitably selecting the power splitting ratio, the time splitting factor, and the weights in the beamforming vector. At the same time, other contributions 
investigated the performance bounds of the secrecy capacity, such as the strictly positive secrecy capacity (SPSC) problem, defined as the probability that the secrecy capacity remains strictly positive.

Another important security metric evaluating the physical layer data secrecy of a transmission scheme is the \emph{\ac{SOP}}, defined as the probability that the instantaneous secrecy capacity drops below a specific threshold value $S$, as defined in the following Eq.~\ref{eq:sop}
\begin{equation}
    \label{eq:sop}
    SOP_n = Pr \left( S_n < S \right) = Pr \left[ \log_2 \left( \frac{1 + SINR_n}{1+SINR_e} \right) < S \right].
\end{equation}

Few papers in the literature focused on an additional security-related metric, which is the \emph{\ac{ESR}}. It is defined as the maximum achievable transmission rate of the legitimate link, not enabling correct decoding of the information on the eavesdropper. As in the following Eq.~\ref{eq:erg_sr}, it can be modeled as the average of the instantaneous secrecy rate jointly considering all the possible channel realizations:

\begin{equation}
\label{eq:erg_sr}
    \begin{split}
        E_n &= E \left[ C_n - C_e \right] = \\
        &= E \bigg\{ \frac{1}{2} \left[ \log_2 \left( 1 + SINR_n \right) - \log_2 \left( 1 + SINR_e \right) \right]^{+} \bigg\},
    \end{split}
\end{equation}
\noindent
where the operator $\left( \cdot \right)^{+}$ refers to the maximum positive value.

Other scientific contributions such as \cite{yu2019_sysj}, \cite{zhao2019_access_2}, and \cite{wang2018_access_3}, focused on the optimization of the \emph{\ac{SEE}} metric, that is a trade-off between the secrecy rate of the communication link and the energy consumption on the EH devices. In practice, energy metrics (such as the power consumption on the transmitter and the consumed energy at the receiver) are coupled with the above security metrics, not to decrease below a given threshold. 

Finally, it is worth noting that the above-introduced metrics could also be treated as constraints of an optimization problem. Thus, many contributions in the literature focus on the optimization of a non-security-related metric, while minimum constraints on the secrecy rate or maximum constraint on the secrecy outage probability are included in the optimization problem to still guarantee data secrecy at the physical layer.

\subsection{Relaying Models and Strategies}
\label{sec:relay}

While the above considerations are valid for single-hop EH networks, many contributions in the literature also investigated multi-hop scenarios, where either the receivers forward the information to intermediate nodes in the network or there is not a direct link between the source and the intended destination. In these cases, dedicated relaying nodes are introduced to accomplish communication tasks.

To enable energy harvesting and information forwarding at the same time, two relaying strategies are mainly discussed in the literature, namely \emph{\ac{DF}} and \emph{\ac{AF}}.

In the DF mode, after the reception of the signal, the relay decodes and re-transmits it as a new signal. In the AF mode, instead, without decoding the signal, the relay amplifies it using a power factor depending on the amount of harvested energy. It is worth noting that, as discussed by the authors in~\cite{Nasir2013}, either the TS or the PS hardware architectures can be adopted on the relay for EH and information processing, leading to different SNR, throughput, and security considerations.

While most of the scientific contributions assumed trusted relays, few of them introduced untrusted relays and provide strategies to protect the information from \emph{honest-but-curious} relays, achieving their forwarding tasks but also interested in stealing the information. More details on this adversary model are provided in Section~\ref{sec:relay_adv}.

\subsection{Artificial Noise}
\label{sec:jamming}

As discussed in the previous sections, the security of the legitimate source-destination communication link, evaluated through metrics such as the secrecy rate and the secrecy capacity (see Sec.~\ref{sec:sec_obj}), is strictly dependent on the difference between the SINR at the legitimate destination and the SINR experienced by the eavesdropper. Thus, an effective strategy to further reduce the SINR at the eavesdroppers is to deploy a \emph{friendly jammer}. A friendly jammer emits artificial noise on the communication channel in a way that its effect is almost nullified at the legitimate receiver, while it is amplified on any other device in the network, leading to an overall decrease of the SINR.

Artificial noise can be introduced in the network by different entities, i.e., a helper node and the destination. When artificial noise is introduced by the helper, the scenario typically involves a dedicated jammer device, continuously emitting noise on the channel. Other contributions couple the roles of the relay and the jammer, selecting helper devices in the set of relays that are not used for signal relaying at a particular time. 

At the same time, artificial noise can be also introduced by the legitimate destination when the receiving device is equipped with multiple antennas.  In this case, while one or more of the antennas are receiving the signal, the remaining ones emit artificial noise, summing up (in voltage) to one of the legitimate signals. On the one hand, the signal available at the receiver results corrupted by the noise. On the other hand, given that the receiving node is aware of the samples of the noise, it can cancel out (by subtraction) the intentional interference and confidentially recover the original signal. Several schemes employing friendly jamming are discussed in the sections below.

\subsection{External Adversaries}
\label{sec:ext_adv}

We define an \emph{external} adversary as a device that does not take part in the regular communication patterns in the network, thus remaining \emph{anonymous} to the network devices. This scenario is typical of any wireless network scenario, including \ac{WSN} and \ac{IoT} networks, where an external eavesdropper simply needs a receiving antenna tuned on the same frequency of the legitimate communication channel to successfully decode the information --- assuming no cryptography techniques are deployed. 

The distinguishing feature of these works is the unavailability of any information about the adversary. From the model perspective, this assumption translates in the general unavailability of the \ac{CSI} experienced by the eavesdropper communication links, further complicating the design of an optimally-secure solution. Indeed, with no information about the adversary and the quality of the channel it experiences, it is possible to work only on the main communication links, either minimizing the probability that the information is correctly decoded by other devices in the network or maximizing the secrecy rate of the main communication links. 

A classification of the scientific contributions assuming external adversaries is provided in Table~\ref{tab:secrecy_ext}, considering the features and capabilities of the adversaries. It is worth noting that all the scientific contributions discussed hereby assumed completely-passive eavesdroppers, that do not transmit anything on the RF spectrum.

\begin{table*}[htbp]
\centering
\caption{Qualitative Comparison among Approaches investigating PHY-layer Data Secrecy in EH Networks confronting external eavesdroppers.}
\begin{tabular}{|P{1.2cm}|P{1.4cm}|P{1.6cm}|P{1.2cm}|P{1.4cm}|P{1.3cm}|P{1.2cm}|P{1.5cm}|P{1.3cm}|P{1.8cm}|}
\hline
\textbf{Adversary} &\textbf{CSI Availability} & \textbf{Adversarial Distribution} &   \textbf{Collusion} & \textbf{Adversarial Receiving Antennas} & \textbf{Adversarial Antenna Type} & \textbf{Adversary Mobility}  & \textbf{Distance Considerations} & \textbf{Scheme} & \textbf{Security Treatment}\\
\hline \hline

External & \xmark & Single & \xmark & Single & Omni-directional & \xmark & \xmark & \cite{pan2017_tcomm}, \cite{wang2018_access}, \cite{mobini2019_tifs} & Minimum \ac{SOP} \\
\cline{9-10}
& & & & & & & & \cite{elshafie2019_commlett} & Constraint, Minimum \ac{SOP} \\
\cline{9-10}
& & & & & & & & \cite{elshafie2017_cl}, \cite{wang2018_tcomm}, \cite{hong2019_access} & Average Secrecy Rate \\
\cline{9-10}
& & & & & & & & \cite{yin2018_access} & SOP, ESR \\
\cline{9-10}
& & & & & & & & \cite{hu2019_commlett} & SOP, SRM \\
\cline{9-10}
& & & & & & & &
\cite{sharma2019_phys}, \cite{wang2019_access} & Secrecy Capacity Maximization \\
\cline{9-10}
& & & & & & & &
\cite{Zhang2016_access} &  SPSC Bounds\\
\cline{9-10}
& & & & & & & & \cite{Chen2016_tvt} & Secrecy Outage Capacity \\
\cline{9-10}
& & & & & & & & \cite{tang2019_tifs} & SOP, Secrecy Rate \\
\cline{9-10}
& & & & & & & & \cite{zhao2020_tvt} & Multiple Users Secrecy Rate \\
\cline{9-10}
& & & & & & & & \cite{wang2020_tcomm2} & Secrecy Rate \\
\hline

External & Imperfect & Single & \xmark & Single & Omni-Directional & \xmark & \xmark & \cite{zhang2015_tc}, \cite{guo2019_iotj} &  \ac{SRM}\\
\cline{9-10}
& & & & & & & & \cite{Bi2016_jstsp} & SOP and SPSC probability\\
\cline{9-10}
& & & & & & & & \cite{zhang2018_tgcn} & SOP \\
\hline

External & \xmark & Single & \xmark & Multiple & Omni-Directional & \xmark & \xmark & \cite{biason2016_jsac} &  \ac{SRM}\\
\cline{9-10}
& & & & & & & & \cite{boshkovska2018_tcomm} & Constraint, Minimum Secrecy Rate \\
\cline{9-10}
& & & & & & & &\cite{liu2018_tvt} & SOP Minimization \\
\hline

External & \xmark & Multiple & \xmark & Single & Omni-Directional & \xmark & \xmark & \cite{Ng2014}, \cite{salem2016_tc}, \cite{lee2019_tifs} & Secrecy Capacity Definition and Maximization \\
\cline{9-10}
& & & & & & & & \cite{nguyen2016_access} & \ac{SOP} \\
\cline{9-10}
 &  &  &  &  & &  &  & \cite{moon2017_tcn}, \cite{niu2019_sysj} &  \ac{SRM}, \ac{SOP} Minimization \\ 
\cline{9-10}
 &  &  &  &  & &  &  & \cite{lu2019_jsac} & Constraint, Minimum SOP \\
 \cline{9-10}
 &  &  &  &  & &  &  & \cite{hu2019_wocc} & Secrecy Capacity \\ 
\hline

External & \xmark & Multiple & \xmark & Single & Directional & \xmark & \xmark & \cite{cao2019_wcl} & SRM, Ergodic Secrecy Capacity\\
\hline

External & Imperfect & Multiple & \checkmark & Single & Omni-Directional & \xmark & \xmark & \cite{li2019_access} & SOP \\
\hline

External & Imperfect & Multiple & \xmark & Multiple & Omni-Directional & \xmark & \xmark &  \cite{niu2017_commlett}, \cite{khandaker2019_tcomm} & Constraint, Minimum Secrecy Rate \\
\cline{9-10}
 &  &  &  &  & &  &  & \cite{li2018_tvt} & SOP \\
\cline{9-10}
 &  &  &  &  & &  &  & \cite{zhao2019_access_1} & Secrecy Rate \\
\hline
External & Imperfect & Multiple & \xmark & Single & Omni-Directional & \xmark & Imperfect Locations & \cite{wang2020_tc} & Secrecy Rate \\
\hline

External & \xmark & Multiple & \xmark & Single & Directional & \xmark & HPPP & \cite{sun2019_access} & Secrecy Rate \\
\cline{9-10}
 &  &  &  &  &  &  &  & \cite{sun2019_commlett} & Lower-bound SRM\\
\hline

\end{tabular}
\\
\label{tab:secrecy_ext}
\end{table*}

We notice that, overall, very different adversarial models are considered in the literature.

\textbf{Single Single-Antenna External Eavesdropper.} The simplest adversary model is a single external eavesdropper, equipped with a single omnidirectional antenna, such as the one assumed in~\cite{pan2017_tcomm}---\nocite{wang2018_access, mobini2019_tifs, elshafie2019_commlett, elshafie2017_cl, wang2018_tcomm, yin2018_access, hu2019_commlett, sharma2019_phys, wang2019_access, Zhang2016_access, Chen2016_tvt}\cite{tang2019_tifs}.

\noindent
To provide a few examples, the authors in~\cite{pan2017_tcomm} contextualized the concepts of RF-SWIPT in \ac{VLC} systems, studying the energy harvesting performance when visible-light signals are used as a source. Assuming a single-antenna eavesdropper and a random location of the eavesdropper to the receiver, they derived the exact analytical expressions of the asymptotic \ac{SOP}, by using the stochastic geometry method, and they evaluated the overall secrecy rate via simulations. Assuming the same adversarial model, the authors in~\cite{mobini2019_tifs} provided a significant contribution to the definition of the secrecy rate that can be achieved via cooperative jamming strategies. They showed that the deployment of a full-duplex jammer node can improve significantly the secrecy rate, and such improvement is highly dependent on the time split ratio for EH activities, the features of the channel, and the topology of the scenario. The authors in~\cite{wang2018_tcomm} focused on buffer-aided networks, where the messages can be temporarily stored on the relays for several slots, waiting to select the transmission link that maximizes the communication secrecy. Assuming either an offline or an online knowledge of the main communication channels, the authors developed a protocol that can jointly optimize the power allocation at the transmitter and the secrecy rate at the receiver, considering the relay transmit power optimization and link selection strategies. We also highlight the work by the authors in~\cite{Chen2016_tvt}, wherein a constrained relay is considered. This relay is equipped with a large number of input and output antennas, according to the Massive-\ac{MIMO} technology. The authors unveiled explicit closed-form expressions for the secrecy outage capacity of a link where the above-described relay is deployed, both assuming the \acl{AF} and the \acl{DF} relaying mode. 

While the previous contributions assumed no knowledge and previous information about the adversarial channel, the authors in~\cite{zhang2015_tc}---\nocite{guo2019_iotj, Bi2016_jstsp}\cite{zhang2018_tgcn} worked with the \emph{Imperfect CSI} assumption, still considering a single adversary with a single antenna. In these settings, they evaluated the impact of such an assumption on the achievable security metrics. 

\noindent
For instance, the authors in \cite{zhang2015_tc} considered a generic \ac{MISO} scenario. However, differently from previous works, they maximize the worst-case secrecy rate of the communication link by decoupling the optimization problem into three unique optimization problems. Using such a strategy, they found a locally-optimum solution utilizing the alternating optimization algorithm.
In the context of a multi-relay network, the authors in~\cite{guo2019_iotj} introduced a relay selection scheme where the relays that can not decode the signal assume the role of friendly jammers, further improving the physical layer security of the scheme compared to regular beamforming schemes. Differently, the authors in \cite{Bi2016_jstsp} focused on the design of a novel jammer, working accordingly to the \emph{Accumulate-and-Jam} logic. This jammer, being energy-constrained, stores energy locally and then uses such energy to perform cooperative jamming. It is worth noting that, differently from other approaches, here the focus is mainly on the communication link between the source of a message and the jammer, where the jammer is intended as a destination. The authors investigated both the design logic of the jammer and its performance in a wiretap channel, considering a single-antenna eavesdropper. Finally, they show the benefits of such design compared to a regular Half-Duplex approach.

\textbf{Single Multi-Antenna External Eavesdropper}. While the above contributions assumed a single adversarial device equipped with a single antenna, other works, such as \cite{biason2016_jsac}---\nocite{boshkovska2018_tcomm}\cite{liu2018_tvt} provided similar studies assuming a more powerful external adversary, characterized by multiple receiving antennas. For instance, the authors in~\cite{boshkovska2018_tcomm} analyzed the effect of non-idealities such as residual hardware impairments in EH devices, CSI imperfections, and non-linearity of EH circuits on the security and energy performance of a SWIPT EH network. In this challenging system model, they considered the secrecy rate of the system toward a multiple-antenna eavesdropper, and they minimized the total power consumption while, at the same time, guaranteeing the minimum required secrecy rate of the main communication links. 

\textbf{Multiple Non-Colluding Single-Antenna External Eavesdroppers}. A slightly different adversarial model is the one where there are multiple adversaries, each equipped with a single omnidirectional receiving antenna. It is worth noting that this adversarial model could be either more powerful or similar to the one where there is a single adversary equipped with multiple antennas, based on the possibility that the adversaries collude. Indeed, if the adversaries do not collude, i.e., do not share information, this model has the same strength and modeling than the single adversary - single antenna attacker model. Instead, when the adversaries can share information, additional statistic analysis and processing tools can be used, based on the difference in the location where the raw signals are acquired.

\noindent
The contributions in \cite{Ng2014}\nocite{salem2016_tc, moon2017_tcn, lee2019_tifs, nguyen2016_access, niu2019_sysj}---\cite{lu2019_jsac} assumed non-colluding adversaries, and investigated the secrecy and energy performance of RF EH networks in this adversarial models. 
For instance, the authors in \cite{Ng2014} considered a \ac{MISO} system, where a single source node, equipped with multiple antennas, would like to transmit a signal to a specific destination node. The authors formulated an optimization problem by defining the channel secrecy capacity as the system performance metric to be optimized, and they used the semi-definite relaxation mathematical tool to solve the optimization problem. Their general finding is that increased values of the SINR lead to increased secrecy capacity, while the proposed resolution strategy helps to reduce the overall transmit power to achieve a given level of secrecy capacity, compared to traditional approaches. Assuming the same adversarial model, the authors in \cite{salem2016_tc} considered the effect on the secrecy rate of destination-assisted jamming, as discussed in Section~\ref{sec:jamming}. Assuming a relay-based network working in the \ac{AF} mode, they organized the transmission of the secret signal in two stages. In the first one, when the source node transmits the signal to the relay, the destination emits artificial noise to both protect the secrecy of the signal and power the relay node. Then, in the second phase, the destination receives the message from the relay and cancels out its own signal, recovering the intended message from the source node. The authors provided a thorough investigation of the secrecy capacity in this scenario, assuming either TS-based or PS-based hardware architecture, and considering the impact of several system parameters, including the time dedicated to EH, the power splitting ratio, the noise, and the location of the participating entities. Their study revealed that PS generally outperforms TS when all the other parameters do not change. Besides, they analytically proved that the secrecy capacity of the system can be maximized in several different ways, including the increase of the number of relays, the noise power, and the distance of the eavesdropper to the main communication links.

\noindent
The performance in the presence of a dedicated friendly jammer device are investigated by the authors in \cite{moon2017_tcn}, assuming both the knowledge and the absence of the \ac{CSI} of the eavesdroppers. Assuming a \ac{MISO} scenario, the authors investigated the SRM and the SOP minimization in a WPCN, where the jammer is powered via EH and follows a TS logic. They also extend the results to a MIMO scenario, demonstrating that their solutions are even more successful in protecting data at the physical layer when more users and more eavesdroppers are considered. 

\noindent
A novel two-way relaying WPCN scenario has been considered by the authors in~\cite{lee2019_tifs}, where a source node eliminates its own component from the signal received by the relay, through self-cancellation techniques. In this novel setting, the authors derived optimally secure protocols based on power-splitting and time-splitting, able to maximize the secrecy capacity of the link assuming multiple non-colluding eavesdroppers.


\textbf{Multiple Non-Colluding Multi-Antenna External Eavesdroppers}. Multiple non-colluding eavesdroppers equipped with multiple antennas are considered by the contributions in~\cite{niu2017_commlett}---\nocite{khandaker2019_tcomm, li2018_tvt}\cite{zhao2019_access_1}.
For instance, the authors in~\cite{khandaker2019_tcomm} jointly used constructive interference and artificial noise to achieve secure communications in SWIPT systems. Assuming the sample case of an 8-PSK modulation, they showed that constructive interference pre-coding schemes can achieve significant power savings compared to interference management approaches, while also getting rid of EH outages that characterize legacy approaches. While the main focus on this contribution is on the power management part, the secrecy rate of the link is considered as a constraint of the problem, in a way to decade below a given acceptable threshold.

\textbf{Multiple Colluding External Eavesdroppers}. Multiple colluding external adversaries have been considered recently, only by the authors in~\cite{li2019_access}. In this context, the authors first verified that colluding capabilities improve the performance of the adversary, thanks to the sharing of information. In addition, they demonstrate that a trade-off between security and reliability exists. Indeed, when more relaxed constraints on the secrecy outage probability are imposed, higher SINR values can be achieved, improving the reliability of the communication link.

\textbf{Directional Antennas.} Other significant contributions are the works in \cite{cao2019_wcl}---\nocite{sun2019_access}\cite{sun2019_commlett}, where multiple non-colluding adversaries are equipped with directional antennas. In particular, while omnidirectional antennas are characterized by almost an equal receiving range in all directions, directional antennas can boost the receiving range along a particular direction. From the security perspective, this is an additional powerful feature, as it can help to eliminate interferences caused by other RF activities, while focusing on a given target device or communication link. 

\noindent
\textcolor{black}{Specifically, the authors in~\cite{sun2019_commlett} focused on a mmWave IoT relay network, where the relays are mobile \acp{UAV}. A unique assumption of this work is the inclusion of a statistical distribution of the distance of the eavesdroppers from the main communication link, following an independent \ac{HPPP} distribution with density $\lambda$ on the ground. The inclusion of directional antennas is modeled via the usage of a 3D antenna gain model, to approximate the
antenna gain pattern of all nodes, as described in~\cite{zhu2018_jsac}. The authors analyzed the impact of the density and the distribution of the adversaries around the target device, deriving lower bounds on the average secrecy rate relatively to the transmit power, the power splitting ratio and the deployment of the UAVs.} 
\textcolor{black}{In~\cite{sun2019_access}, the same authors of the above cited paper, were among the first ones to investigate the secrecy performance of mmWave SWIPT UAV-based relay systems. An important difference characterizing this work is in the model of the air-to-ground channels, that are considered as Nakagami-m small-scale fading. Under these new assumptions, the authors computed the expressions of the achievable secrecy rate, assuming both Detect-and-Forward (DF) and Amplify-and-Forward (AF) schemes at the relay, and they investigated the relationship between all the metrics of the system.}

\textcolor{black}{The usage of \acp{UAV} as elements of the network infrastructure can be found also in the contribution by the authors in~\cite{wang2020_tc}. A distinctive feature of this work is in the application of EH capabilities on the UAVs, essentially for information decoding and relay, while an onboard battery supplies energy for the flight control of the UAVs. The authors assume multiple non-colluding terrestrial eavesdroppers, trying to infer on the data transmitted by a source, and relayed by the UAVs to the intended destination. In this scenario, they derived the secrecy rate of the communication link, considering as a new factor of the optimization problem the relative position of the UAV toward the source and the destination. UAVs are also used as transmitters in the contribution by the authors in~\cite{wang2020_tcomm2}, in combination with the \ac{NOMA} technique to achieve secure transmissions against a single single-antenna eavesdropper. To further improve the secrecy of the communication, the authors use artificial friendly jamming, to further decrease the chances for the eavesdropper to fully decode the data delivered by the UAV. Considering a non-linear EH model, the authors derived the secrecy rate of the communication link and maximize jointly the throughput of the communication link, the TS ratio, and the PS ratio.
}

\noindent
Finally, we highlight that while these valuable contributions brought the novelty of directional antennas in the RF EH physical-layer security area, they did not explore collusion capabilities, still posing severe constraints on the capabilities of the adversary.


\subsection{Internal Adversaries}
\label{sec:int_adv}

We define an \emph{internal} adversary as a device that is a legitimate member of the network, actively taking part in the network activities by transmitting and receiving information. In the physical-layer security literature for RF EH networks, such an adversarial model is often referred to as an \emph{active eavesdropper}, being it characterized both by active legitimate features (authorized transmissions/reception operations) and by unauthorized passive features (i.e., unauthorized passive eavesdropping of the communications)~\cite{do2019_sensors}.

While this scenario could be typical of any wireless network, it is particularly adopted in the context of \aclp{CRN}. Specifically, \acp{CRN} involve both primary users and secondary users. Primary users are express-class licensed devices that pay a fee to experience a minimum guaranteed level of \ac{QoS} for their traffic. Secondary users, often referred to as \emph{cognitive users}, are non-licensed users that are allowed to use the same spectrum of primary users provided that their RF operations do not interfere with them or do not cause excessive degradation of the service offered to primary users. Overall, a CRN is classified based on the paradigm used by secondary users to access the spectrum, and it can be either interweave, underlay, or overlay. While the underlying logic and techniques used by \acp{CRN} are out of the scope of this contribution, the most interesting feature of \acp{CRN} within the scope of this survey is the assumed knowledge of the presence of eavesdropping adversaries, often identified in the secondary users, interested in decoding RF activities of the primary users.

\noindent
Given that these devices are part of the network and actively participate in the communications, the \ac{CSI} experienced by these devices can be assumed as well-known or, at least, statistically available on the source of the confidential communication. Thus, the transmission and the security of the communication link can be optimized by degrading its perceived quality on the eavesdroppers, boosting its secrecy rate.

A comprehensive classification of the scientific contributions assuming internal adversaries is provided in Table~\ref{tab:secrecy_int}, with reference to the features and capabilities of the adversaries.
\begin{table*}[htbp]
\centering
\caption{Qualitative comparison among approaches investigating PHY-layer data secrecy in EH networks confronting internal adversaries.}
\begin{tabular}{|P{1.1cm}|P{1.38cm}|P{1.5cm}|P{1.1cm}|P{1.4cm}|P{1.3cm}|P{1.2cm}|P{1.5cm}|P{1.5cm}|P{1.8cm}|}
\hline
\textbf{Adversary} &\textbf{CSI Availability} & \textbf{Adversarial Distribution} &   \textbf{Collusion} & \textbf{Adversarial Receiving Antennas} & \textbf{Adversarial Antenna Type} & \textbf{Adversary Mobility}  & \textbf{Distance Considerations} & \textbf{Scheme} & \textbf{Security Treatment}\\
\hline \hline

Internal & \checkmark & Single & \xmark & Single & Omni-Directional & \xmark & \xmark & \cite{Li2014_TVT}, \cite{son2015_iet}, \cite{shi2015_twc}, \cite{feng2015_tvt}, \cite{elshafie2016_wcl}, \cite{xing2014_globecom}, \cite{elshafie2016_wcl_2}, \cite{mobini2017}, \cite{Khandaker2017_twc}, \cite{zhang2018_tvt}, \cite{deng2019_iotj}, \cite{do2019_sensors} &  \ac{SRM}\\
\cline{9-10}
& & & & & & & & \cite{xiang2018_tifs} & SRM, Constraint SOP \\
\cline{9-10}
& & & & & & & & \cite{zhang2016_tifs} & Constraint, Minimum Secrecy Rate \\
\cline{9-10}
& & & & & & & & \cite{xing2016_tvt}, \cite{Hoang2017}, \cite{lei2017_tgcn}, \cite{wang2018_access_2}, \cite{maji2018_ietcomm}, \cite{zhang2019_wc}, \cite{nguyen2019_ietcomm} & \ac{SOP} Minimization \\
\cline{9-10}
& & & & & & & & \cite{ghosh2017_tccn}, \cite{lee2018_iotj} & Secrecy Capacity Definition and Maximization \\
\cline{9-10}
& & & & & & & & \cite{jameel2018_comlett} & Eavesdropping Probability \\
\cline{9-10}
& & & & & & & &
\cite{wu2019_phy}, \cite{zhao2020_sysj} & Secrecy Rate - Harvesting Ratio Trade-Off\\
\cline{9-10}
& & & & & & & & \cite{ren2020_tvt} & Secrecy Throughput \\

\hline
Internal & Imperfect & Single & \xmark & Single & Omni-Directional & \xmark & \xmark & \cite{Ng2016_tvt} &   Constraint, Minimum SOP\\
\hline

Internal & \checkmark & Single & \xmark & Multiple & Omni-Directional & \xmark & \xmark & \cite{nasir2017_tsp}, \cite{alageli2019_jsac} & Minimum \ac{SRM} \\
\cline{9-10}
 &  &  &  &  & &  &  & \cite{alageli2019_jsac} & ESR \\
\cline{9-10}
 &  &  &  &  & &  &  & \cite{wang2019_infocom} & Constraint, Minimum Secrecy Rate \\
\hline

Internal & \xmark & Single & \xmark & Single & Omni-Directional & \xmark & \xmark & \cite{ji2020_tccn} & SOP, SRM \\
\hline

Internal & N/A & Single & \xmark & Multiple & Omni-Directional & \xmark & \xmark & \cite{saber2019_sysj} & Avg. Secrecy Capacity, Lower Bound SOP\\
\hline

Internal & \checkmark & Multiple & \xmark & Single & Omni-Directional & \xmark & \xmark & \cite{le2015_commlett}, \cite{Huang2017_access} &  Constraint, Minimum Eavesdropper SINR \\
\cline{9-10}
& & & & & & & & \cite{gao2016_spl}, \cite{tang2018_adhoc} & Constraint, Maximum \ac{SOP} \\
\cline{9-10}
& & & & & & & & \cite{jiang2016_cl}, \cite{boshkovska2017_wcncw} & Constraint, Minimum Eavesdropper Secrecy Rate \\
\cline{9-10}
& & & & & & & & \cite{vo2017}, \cite{vo2018_access_1}, \cite{bi2018_tvt}, \cite{yan2018_tvt}, \cite{nguyen2019_mca} & SOP \\
\cline{9-10}
& & & & & & & & \cite{alageli2018_tvt}, \cite{ma2019_tcomm} & SRM \\
\hline

Internal & Imperfect & Multiple & \xmark & Single & Omni-Directional & \xmark & \xmark & \cite{zhou2018_jsac} &   SOP\\ 
\cline{9-10}
 &  &  &  &  & &  &  & \cite{ni2018_access} & Constraint, SOP \\
\hline

Internal & \xmark & Multiple & \xmark & Multiple & Omni-Directional & \xmark & \xmark & \cite{wang2018_access_3}, \cite{yu2019_sysj}, \cite{zhao2019_access_2} &  Constraint, Minimum Secrecy Rate  \\
\cline{9-10}
 &  &  &  &  & &  &  & \cite{li2018_tvt} & SOP \\
\hline

Internal & \checkmark & Multiple & \checkmark & Single & Omni-Directional & \xmark & \xmark & \cite{hieu2018_sensorsJ} &  \ac{SOP}  \\
\hline

Internal & Imperfect & Multiple & \checkmark & Single & Omni-Directional & \xmark & \xmark & \cite{khandaker2015_tifs} &  \ac{SRM}  \\
\hline

Internal & \checkmark & Multiple & \checkmark & Multiple & Omni-Directional & \xmark & \xmark & \cite{niu2017_wcl}, \cite{khandaker2017_tifs}, \cite{chen2018_access} &  Constraint, Maximum Outage Secrecy Rate  \\
\hline
\end{tabular}
\\
\label{tab:secrecy_int}
\end{table*}

\textbf{Single Single-Antenna Internal Eavesdropper.} The most common scenario investigated in the literature assumes a single internal adversary, whose CSI is well-known, equipped with a single omni-directional antenna, as in the contributions by the authors in \cite{Li2014_TVT}---\nocite{son2015_iet, shi2015_twc, feng2015_tvt, elshafie2016_wcl, xing2014_globecom, elshafie2016_wcl_2, mobini2017, Khandaker2017_twc, zhang2018_tvt, deng2019_iotj, do2019_sensors, xiang2018_tifs, zhang2016_tifs, xing2016_tvt, Hoang2017, lei2017_tgcn, wang2018_access_2, maji2018_ietcomm, zhang2019_wc, nguyen2019_ietcomm, ghosh2017_tccn, lee2018_iotj, jameel2018_comlett}\cite{wu2019_phy}.

\noindent
To provide a few examples, the authors in \cite{xing2014_globecom} were the first to consider a system with a single source node and one destination node, where the information transfer is aided by a network of cooperative relays. In this scenario, all the relays work according to the EH-TS logic, and they are partitioned in two subsets. Some of them only forward packets to the destination, while the remaining ones, also referred to as \emph{helpers}, act as friendly jammers, injecting random noise. This \emph{jamming} operation is performed smartly, to minimally affect the \ac{SINR} of the signal on the intended destination, while degrading the most the \ac{SINR} at any other location.  At the same time, the scenario includes a single-antenna eavesdropper, located far from the main communication links. In the paper, the authors obtained via the semi-definite relaxation mathematical tool the configuration of the beamforming matrix that can achieve the maximization of the secrecy rate of the main communication link, by also controlling the power consumption on the relays. 
Considering the same system scenario and adversary model, the authors in \cite{Hoang2017} extended the analysis, and they found that the secrecy of the data can be enhanced either by increasing the number of relays, or by increasing the \ac{SINR} at the destination, or by balancing the TS ratio between energy harvesting and information decoding. 

\noindent
In the same network model assumed by the authors in \cite{Hoang2017}, the authors in \cite{elshafie2016_wcl} modeled the batteries of the destination node as a queuing system, and they proposed a scheme to maximize the secrecy rate optimizing the transmission time of the source node. At the same time, the artificial noise emission by the jammer is optimized to cancel at the destination's location, while being effective at other locations.
Assuming a similar scenario than \cite{Hoang2017} and \cite{xing2014_globecom}, the authors in \cite{Li2014_TVT} considered an SR hardware design, and a single relay, only working according to the TS-AF scheme. The authors assumed that the CSI of the whole network is known and, under this assumption, they proposed an iterative algorithm rooted in the Constrained Concave Convex Procedure (CCCP) mathematical tool. Their proposed algorithm provides a locally-optimum solution, maximizing the secrecy rate of the link, simultaneously fulfilling the constraints on the overall transmission power and harvested energy on the receiver.
Still assuming a SISO system but with a \ac{PS} hardware architecture, the authors in \cite{son2015_iet} first introduced the concept of the active eavesdropper. They focused on the relaying logic, evaluating the secrecy capacity of the communication link when either the DF or the AF strategy is used by the EH relays. Via Monte-Carlo simulations, they demonstrated the DF strategy to be more \emph{confidential} than the AF strategy, and that both the strategies have the potential to improve the secrecy rate when an increased number of relays are deployed, and when the relays are closer to the destination than the eavesdropper.

\noindent
A \ac{MIMO} system has been considered by the authors in \cite{shi2015_twc}, where the receivers are based on the \ac{SR} hardware architecture. Initially, without including any relay nor jamming activity in the system model, they considered two specific scenarios (single-stream and full-stream), and for each of them they point out globally optimum solutions based on the semi-definite relaxation mathematical tool. Then, for an expanded scenario including relays producing artificial noise via jamming, they discuss the solution of an equivalent problem based on the Inexact Block Coordinate Descent (IBCD) algorithm, and they evaluate its performance in maximizing the secrecy rate of the communication link. In the same scenario, the authors in \cite{feng2015_tvt} considered a receiver equipped with only a single antenna, and modeled the channel as slow-fading. They formulated the optimization problem to maximize the worst-case secrecy, assuming the overall transmission power and the amount of harvested energy as the constraints of the optimization problem.

\noindent
Despite many contributions considered the involvement of relays in the system, the reason motivating network devices to behave as nodes has received only minimal attention by the community. In this context, a pioneering contribution has been provided by the authors in~\cite{Khandaker2017_twc}, where the relays are selected by the transmitter from a set of devices, and they are following remunerated in the form of minimum provided energy, that can be harvested and used by the relay to fulfill their own tasks in the network. To ensure fair participation of the relays, and to motivate them to reveal trustful information, the authors developed a relay selection strategy enforcing their honest behavior, while dishonest ones are penalized by decreasing their trust degree.

\noindent
The secrecy rate at the physical layer has been investigated also by the authors in~\cite{zhang2018_tvt}, considering a network using the \ac{OFDM} scheme. To improve the achievable secrecy rate, the authors deployed a jammer that is cooperative with the legitimate devices. The jammer works accordingly to the harvest-and-jam logic, gaining energy from the information message in the first phase of the information transmission, and jamming the message at the eavesdropper in the second phase. Differently from previous contributions assuming this scenario, following a logic similar to the previous contribution in~\cite{xu2014_tsp}. the authors considered two different kinds of receivers, namely Type I and Type II, equipped or not with the capabilities of removing the noise on the reception side, respectively. Thus, they derive \ac{SRM} definition in this scenario with the above constraints, optimizing simultaneously the source transmit power and the jammer signal over the OFDM sub-carriers.

\noindent
The authors in \cite{xing2016_tvt} used artificial noise at the transmitter, by splitting the transmission power into two parts: one for information transmission and one for noise emission. A key pre-distribution scheme is assumed between the transmitter and the receiver, in a way that the noise could be canceled out at the receiver but not at the transmitter. 

\noindent
The authors in \cite{zhang2016_tifs} integrated SWIPT concepts within cellular networks based on the \ac{OFDMA} modulation scheme. Specifically, assuming the secrecy rate as a constraint of the problem, they provide two sub-optimal solutions for the maximization of the harvested power, by efficiently allocating the power on different sub-carriers.
The authors in~\cite{lei2017_tgcn} derived the \ac{SOP} of two competing antenna-selection schemes, namely the optimal antenna selection (OAS) scheme and suboptimal antenna selection (SAS). They compared the performance of these schemes against traditional transmission schemes via simulations, verifying that OAS outperforms SAS when RF constraints are considered. 
Differently from other contributions, the authors in~\cite{xiang2018_tifs} investigated also power optimization problems on the transmission side. Specifically, they evaluated the best power control policy on the transmitter, achieving the secrecy rate maximization on the receiver. To this aim, they derive a relationship between the secrecy rate and the SNR at the receiver, developing an algorithm that reduces at minimum the complexity and the computational burden on the transmitting device. Moreover, their evaluation shows that a high transmission power does not guarantee, at the same time, minimum secrecy rate and energy harvesting requirements.
An interesting contribution is provided also by the authors in~\cite{lee2018_iotj}. In this contribution, the authors proposed two relaying schemes based on power splitting and time-splitting (namely, PSR and TSR), where the relays can tune the ratios in a way to trade-off computational and harvesting processes. Interestingly, they provide results showing that the channel information at the eavesdroppers does not affect these metrics (note that the usefulness of the CSI is not discussed in the derivation of the maximum secrecy rate of the system). Finally, they show that the PSR scheme outperforms the TSR when secrecy capacity is considered.

\noindent
It is worth also mentioning the work by the authors in~\cite{do2019_sensors}, where the adversary is a jammer, besides being characterized by the standard eavesdropping capabilities.


\noindent
The \emph{Imperfect CSI} assumption is used in \cite{Ng2016_tvt}, still considering a single single-antenna eavesdropper scenario. The authors assumed the secondary users to be energy-constrained devices, requiring power from the primary users to operate in the network. Given that these devices are constrained, the authors assume that they do not have enough capabilities to perform energy harvesting and data decoding at the same time; thus, a \ac{TS} architecture is assumed. In this context, the authors designed a multi-objective optimization framework based on a Pareto-optimal resource allocation algorithm, leveraging the weighted Tchebycheff approach, where the maximization of the secrecy rate is one of the objectives of the optimization.

\textbf{Single Multi-Antenna Internal Eavesdropper.} A single internal eavesdropper equipped with multiple antennas is considered by the authors in~\cite{nasir2017_tsp}---\nocite{alageli2019_jsac}\cite{wang2019_infocom}.

\noindent
The contribution by the authors in~\cite{alageli2019_jsac} is particularly interesting from the security perspective, given that they considered not only information eavesdropping but also unauthorized EH activities. This work considered the adversary as a two-antenna active energy harvester, working according to the SR logic, able to legitimately harvest energy via one antenna, and illegitimately and actively eavesdropping the signal using the second antenna. Therefore, during information transmission activities, this node is a regular eavesdropper. At the same time, during EH activities, it is interested in acquiring much energy and recharging its battery more than expected, using hardware circuitry that is more performant than the legitimate devices. The authors formulated asymptotic expressions of the lower bound on the ESR and the average harvested energy, and they used these results to optimize the system design, reducing the power and information leakage based on several system parameters. They also demonstrated the existence of a trade-off between the worst-case \ac{ESR} and the worst-case average for the harvested energy.
It is worth also mentioning the recent work in~\cite{wang2019_infocom}, where the authors experimentally verified their findings through a real proof-of-concept, using off-the-shelf Software Defined Radios and backscatter systems. Note that this is the only work on physical-layer data secrecy that provides experimental tests and verification of the theoretical findings. 
No knowledge about the eavesdropper's channel conditions is also assumed in the work by the authors in~\cite{saber2019_sysj}, where a Nakagami-$m$ channel model is assumed. The context assumed by this work is very specific, as it considers a SIMO RF network integrating SWIPT concepts mixed with a free-space optical communication link. The authors considered a single energy harvesting receiver, potentially behaving as an eavesdropper, and they investigated the average secrecy capacity and the lower bound of the \ac{SOP} taking into account different physical-layer parameters, such as the architecture of the receiver and the deployment of multiple antennas.

\textbf{Multiple Single-Antenna Internal Eavesdroppers.} Multiple non-colluding eavesdroppers equipped with single omni-directional antennas are considered as adversaries in~\cite{le2015_commlett}---\nocite{Huang2017_access, gao2016_spl, tang2018_adhoc, jiang2016_cl, boshkovska2017_wcncw, vo2018_access_1, bi2018_tvt, alageli2018_tvt}\cite{ma2019_tcomm}.
Specifically, the authors in \cite{le2015_commlett} have been the first to integrate EH concepts in the context of cellular networks. Here, the Cellular Network Base Stations are assumed as the transmitters, and the authors investigated the secrecy capacity of the link assuming the receivers are equipped with EH modules (considering an SR hardware architecture). The authors formulated the optimization problem to minimize the overall transmit power of the system, while taking the link secrecy rate as a constraint of the optimization problem, and they used the Semi-Definite Relaxation tool to find a convex form of the problem.

\noindent
The authors in~\cite{bi2018_tvt} proposed a protocol to enhance the secrecy rate of a WPCN, thanks to the deployment of a power beacon. First, the power beacon transmits RF energy to charge a node. Then, it protects the transmission of this device by emitting friendly jamming, thus degrading the SNR at the eavesdroppers. They studied the problem using finite-state Markov Chains, deriving the expressions of the SOP and secrecy rate in this scenario, evaluating also the reliability and security of the communication link. 

\noindent
In the context of overlay \acp{CRN}, the authors in~\cite{yan2018_tvt} proposed and evaluated a dedicated scheme, namely EARTH, where the residual harvested energy is assumed as a constraint of an optimization problem. Assuming as a metric the trade-off between secrecy rate and outage probability, they derived the SOP of the system and compared its performance against standard scheduling methods, such as round-robin, showing superior performance. 
Assuming a MISO scenario, the authors in~\cite{ma2019_tcomm} considered the maximization of the sum logarithmic secrecy rates, by considering energy harvesting constraints. To achieve the objective, they use specific optimization tools, such as the \ac{SDR} and the successive convex centralized beamforming design. They demonstrated that the transmission of single streams does not affect the secrecy rate.

\noindent
\textcolor{black}{Imperfect knowledge of the adversarial CSI is assumed in the works by the authors in~\cite{zhou2018_jsac} and~\cite{ni2018_access}. 
In particular, the authors in~\cite{zhou2018_jsac} provided a complete study on the performance of EH-\ac{NOMA} schemes in \acp{CRN}, by assuming a realistic non-linear EH model. Also in this case, the physical-layer data secrecy is enhanced thanks to the deployment of cooperative jamming techniques. The authors used the \ac{SDR} method to derive closed-form expressions for the problem, and showed that their proposed solution achieves better secrecy rate than the conventional \ac{OFDM} schemes. }

\textbf{Multiple Non-Colluding Multi-Antenna Internal Eavesdroppers.} Multiple non-colluding multi-antenna eavesdroppers are considered in~\cite{yu2019_sysj}---\nocite{zhao2019_access_2}\cite{wang2018_access_3}, and~\cite{li2018_tvt}, all investigating the \ac{SEE} problem. Besides introducing and formalizing the SEE problem, the authors in \cite{yu2019_sysj} investigated a weighted-sum secrecy rate maximization problem, considering at the same time the secrecy rate of the communication link and the total power consumption on the transmitter. They solved the problem using an iterative algorithm based on the Dinkelbach method \cite{zappone2015}, and they validated the effectiveness of such a method via numerical analysis. In the context of overlay \acp{CRN}, the authors in~\cite{li2018_tvt} proposed optimal relay selection (ORS) strategies assuming both the time switching relay (TSR) protocol and power splitting relay (PSR) protocol. They derived exact closed-form expressions related to the \ac{SOP} of primary users and the ergodic rate of secondary users. An important contribution by the authors is the derivation of the SOP of the primary users assuming an infinite number of eavesdroppers and high SNR on the adversarial side.
\noindent

The authors in~\cite{khandaker2015_tifs} considered multiple colluding single-antenna eavesdroppers, where the \emph{maximum SINR receive beamforming} technique is used by the eavesdroppers to aggregate information. The authors formulated the optimization problem to maximize the secrecy rate of the main communication link, considering the total harvested power and the power consumption as constraints of the optimization problem. They found that despite the maximum secrecy rate is constrained by the total harvested power constraint, only a little reduction in the secrecy rate is achieved, especially when the information about the CSI of the eavesdroppers is available.

\textbf{Multiple Colluding Multi-Antenna Internal Eavesdroppers.} Multiple colluding multi-antenna eavesdroppers are considered in~\cite{niu2017_wcl}---\nocite{khandaker2017_tifs}\cite{chen2018_access}, with either imperfect or full CSI knowledge.  In this context, while most of the schemes considered a perfect or imperfect knowledge of the \ac{CSI} at the eavesdroppers, the authors in~\cite{khandaker2017_tifs} considered a realistic scenario, where the CSI of the eavesdroppers is not known by the legitimate entities. Assuming multiple, multi-antenna, possibly-colluding eavesdroppers, the authors formulated a power minimization problem, assuming the maximum achievable secrecy rate as a constraint of the problem. They also found the maximum secrecy rate of the system, assuming as constraints of the problem a maximum transmission power and a minimum outage probability.

\noindent
The authors in~\cite{hieu2018_sensorsJ} investigated the multi-hop multipath wireless network scenario, where all the nodes in the path are energy-harvested devices. They proposed three different protocols for the path selection from a source node to a destination node, to minimize the SOP and maximize the harvested energy, at the same time. Besides considering multiple non-cooperative eavesdroppers, they also include hardware impairments in the model, deriving expressions that closely match real deployments.

\noindent
Finally, we highlight that all of the above approaches considered neither directional antennas nor included considerations regarding the physical distribution of the eavesdroppers in the scenario. At the same time, in line with the results of our analysis regarding external adversaries, mobile adversaries are not considered.

\subsection{Untrusted Relays}
\label{sec:relay_adv}

We already introduced in Section~\ref{sec:relay} and discussed in the two previous subsections the deployment motivations and main working strategies of \emph{trusted} relay nodes. Especially in deep-fading and highly distributed scenarios, dedicated relay nodes provide a crucial service, by forwarding a (possibly amplified) version of the signal from the source toward the destination. 

However, relays can be characterized by malicious behaviors. Indeed, they could be either untrusted devices, deployed by network operators other than the ones operating the EH network, or they can be attacked by malicious adversaries. In both scenarios, an attacker controlling the relay owns a crucial network element, potentially disclosing a significant part of the information. Especially when working with physical-layer security measures only (as in the contributions discussed hereby), no further protection measures can ensure the confidentiality of the information. 
These are the main reasons motivating many studies in RF EH networks, where untrusted relays are considered. The most common model is a \emph{honest-but-curious} relay, still achieving its relaying tasks in the networks, but trying also to stealthily access the information exchanged between the source and the destination. 

A classification of the scientific contributions assuming untrusted relays in RF EH networks is provided in Table~\ref{tab:secrecy_relay}, considering the features and capabilities of the untrusted relay(s).

\begin{table*}[htbp]
\centering
\caption{Qualitative Comparison among Approaches investigating PHY-layer Data Secrecy in EH Networks assuming untrusted relays.}
\begin{tabular}{|P{1.2cm}|P{1.4cm}|P{1.6cm}|P{1.1cm}|P{1.4cm}|P{1.3cm}|P{1.2cm}|P{1.5cm}|P{1.2cm}|P{1.8cm}|}
\hline
\textbf{Adversary} &\textbf{CSI Availability} & \textbf{Adversarial Distribution} &   \textbf{Collusion} & \textbf{Adversarial Receiving Antennas} & \textbf{Adversarial Antenna Type} & \textbf{Adversary Mobility}  & \textbf{Distance Considerations} & \textbf{Scheme} & \textbf{Security Treatment}\\

\hline \hline
Untrusted Relay & \checkmark & Single & \xmark & Single & Omni-Directional & \xmark & \xmark & \cite{kalamkar2017_tvt}, \cite{he2017_ietcomm}, \cite{mamaghani2019} &  \ac{ESR} and \ac{SOP} Definition \\
\cline{9-10}
& & & & & & & & \cite{Shi2019_sensors}, \cite{shi2019_access}, \cite{chen2019_jcn} & SOP and Secrecy Rate \\
\cline{9-10}
& & & & & & & & \cite{yao2018_access} & SRM \\
\hline

Untrusted Relay & Imperfect & Single & \xmark & Single & Omni-Directional & \xmark & \xmark & \cite{lee2020_tifs} & Average Secrecy Rate  \\
 \hline
 
Untrusted Relay & \checkmark & Multiple & \xmark & Single & Omni-Directional & \xmark & \xmark & \cite{vo2018_access_2} &  SOP and Secrecy Rate\\
\cline{9-10}
& & & & & & & & \cite{elshafie2017_access} & \ac{ESR} \\

\hline
\end{tabular}
\\
\label{tab:secrecy_relay}
\end{table*}

\textbf{Single Untrusted Relays.} A single untrusted relay equipped with a single omni-directional antenna is assumed as the adversarial model by the authors in~\cite{kalamkar2017_tvt}---\nocite{he2017_ietcomm, elshafie2017_access, mamaghani2019, Shi2019_sensors, shi2019_access}\cite{chen2019_jcn}.

\noindent
To protect the information from eavesdropping on the relay, the authors in~\cite{kalamkar2017_tvt} described a destination-assisted jamming strategy, where the destination coordinates with the source to transmit a jamming signal in parallel with the reception of the information signal transmitted by the source to the relay. Besides providing benefits in terms of \ac{EH} (the relay is powered by the jamming signal), the destination-assisted jamming also enhances the security of the information at the relay. Indeed, the signal received at the relay is the combination of the jamming and the information, not revealing the main message to the relay. When the relay forwards the original message to the destination, the final receiving node can subtract the noise and recover the original information. 
The authors evaluated the \ac{SOP} and the \ac{ESR} assuming the usage of both the TS and the PS strategy at the relay, and they concluded that extended benefits in terms of physical-layer data secrecy are provided via the use of the PS strategy. Similar advantages in terms of data secrecy can be provided when the relay is closer to the destination than to the source.

\textcolor{black}{
Still assuming a single untrusted relay, the authors in \cite{lee2020_tifs} recently studied the effect of an outdated CSI estimation on the achievable outage performance and secrecy rate of a communication link leveraging an untrusted relay. Similarly to previous work, this contribution involves the use of friendly jamming strategies to hide the message from the relay. However, due to the imperfection of the CSI estimate, this contribution also studied the impact of an imperfect jamming cancellation on the secrecy performance of the communication link. Specifically, this work finds the expression and configuration of several parameters, such as the jamming power ratio, power splitting factor and time-splitting ratio that can optimize the secrecy performance metrics, finding sub-optimal but realistic solutions.
}

\noindent
Despite the authors in~\cite{kalamkar2017_tvt} assumed the use of the DF strategy on the relay, all the other works listed in Tab.~\ref{tab:secrecy_relay} assumed a relay working according to the AF strategy. In general, this latter technique is preferable from the security perspective, as it does not allow the relay to decode the information while in transit toward the destination.  

\noindent
Assuming destination-assisted jamming, the authors in \cite{shi2019_access} derived the closed-form expression of the secrecy rate and the SOP for a MIMO network, where a base station equipped with multiple antennas transmits confidential information toward multi-antenna receivers. They also showed that increasing the SINR on the destination provides enhancement on the secrecy rate, even if the secrecy rate approaches a constant value by increasing the SINR.

\textbf{Multiple Non-Colluding Untrusted Relays.} Multiple non-colluding untrusted relays equipped with a single receiving antenna are considered only in the work by the authors in~\cite{vo2018_access_2} and~\cite{elshafie2017_access}. For instance, the authors in~\cite{elshafie2017_access} delegated the jamming activities to a dedicated friendly jamming device, emitting noise according to a time profile that is known (and thus, cancellable) at the receiver.

\subsection{Lessons Learned}
\label{sec:secrecy_lessons}

The discussion and classification reported in the previous subsections highlighted several take-away messages, summarized in the following.

\textbf{Security and Energy Trade-off.} 
Security and energy are often contrasting objectives. Indeed, when the maximum secrecy rate is required, more energy than the optimal is required from the source, leading to a non-optimal energy efficient configuration. At the same time, minimizing the power consumption on the source or maximizing the harvesting ratio on the receiver always leads to sub-optimal secrecy rate, as well as secrecy outage effects. In this context, several approaches, such as \cite{yu2019_sysj}, \cite{zhao2019_access_2}, and \cite{wang2018_access_3}, introduced hybrid security-energy metrics, trading off between security and energy availability. Considering the overall scarce energy availability of RF EH scenarios, configuring the network in a way to achieve such a trade-off is crucial to guarantee, at the same time, network lifetime, usability, and reliability constraints.

\textbf{Channel State Information (CSI) Awareness.} The availability of the \ac{CSI} on the eavesdropper(s) side is crucial in calibrating the optimality of the solution from the security perspective. When the CSI of the eavesdropper-source link is known, it is possible to maximize the secrecy capacity and to select the beamforming vector and the jamming to minimize the SNR on the adversary. Another common approach is to consider data secrecy or secrecy capacity as a constraint of the optimization problem, assuring that the secrecy rate does not degrade over a minimum threshold or, equivalently, the secrecy outage probability does not exceed a specified upper bound. 

\textcolor{black}{
\textbf{Physical-Layer Security and Data Secrecy issues in Non-Orthogonal Multiple Access (NOMA) networks.} During the last few years, a significant amount of scientific contributions have focused on physical-layer and data secrecy security issues in EH networks working according to the Non-Orthogonal Multiple Access (NOMA) paradigm. Compared to previous technologies based on orthogonal multiple access, such as OFDMA, NOMA schemes allow overcoming issues related to the limited number of orthogonal resources available in the spectrum and the limited number of users that can be served, tolerating interferences and removing them on the user's side thanks to Successive Interference Cancellation (SIC) techniques~\cite{elbayoumi2020_comst}. However, the application of such advanced techniques generally introduce an increase in the cost and complexity of the receivers, drastically impacting the energy budget of such devices. Therefore, one of the key challenges that have attracted the interest of Academia is the trade-off between energy and security, guaranteeing that users can harvest sufficient energy from the received signals, while also guaranteeing minimum thresholds on the security-related metrics, such as the secrecy rate. Other recent works such as~\cite{zhao2020_tvt} have also brought UAVs into the play, improving the SINR on the receivers and managing energy-security trade-offs on the UAVs.}

\textbf{Relationship between SINR and location.} The largest part of the papers discussed in the previous subsections did not specifically consider the physical deployment of the adversary, i.e., its position to the main communication link. This is because the distance eavesdropper-target is mimicked by the SINR experienced at the eavesdropper. Without loss of generality, the above contributions assumed a homogeneous scenario, where high values of eavesdropping SINR correspond to locations of the eavesdropper close to the target communication link, while low values of SINR experienced by the eavesdropper are consistent with a positioning far from the main communication link. However, translating SINR values to a real distance from the target has been not investigated in the literature.

\subsection{Future Directions}
\label{sec:secrecy_future}

Despite the large body of scientific contributions available in the literature, EH security issues still include some unexplored research directions, listed below.

\textbf{Multiple Colluding Directional Antennas.} Few recent works investigated the secrecy rate that can be achieved via physical-layer security schemes when the adversaries use a directional antenna (see~\cite{sun2019_access} and~\cite{sun2019_commlett}). However, to the best of the authors' knowledge, none of these contributions considered the use of multiple directional antennas. Such a scenario could be particularly interesting from the security perspective, especially when the adversary could take advantage of different eavesdropping locations, such as when collusion is considered. In this scenario, the adversary could effectively reject or minimize the impact of the intentional interference caused by other network elements, thus experiencing an SINR very close to the one in the target destination location. 

\textbf{Antenna Identification Mechanisms.} All the approaches discussed in this section investigated physical-layer security from the information-theoretic perspective, looking at the security metrics discussed in Section~\ref{sec:sec_obj}. However, given that these strategies deal with RF signals, further security aspects should be analyzed, such as the possibility to identify single-antenna contributions to the final signal. For instance, in the context of audio signals, the \ac{ICA} method has been already demonstrated to be a successful tool to identify and reconstruct the waveform emitted by a single antenna \cite{caprolu2020_comst}, thanks to the deployment of a network of colluding eavesdropping antennas. To the best of the authors' knowledge, these physical-layer security analyses have not been provided in the literature.

\textcolor{black}{
\textbf{UAVs and Mobile Adversary.} Despite some recent work assuming RF EH devices included mobile devices, such as \acp{UAV}, the adversary (either single or multiple) is always considered as a static network element, and only a single work (e.g., \cite{son2015_iet}) assumed location diversity. We notice that a mobile adversary could explore location diversity techniques to boost its SINR, as well as it could get closer to the target (mobile) adversary to improve its data decoding chances. In addition, multiple adversarial UAVs can be even more capable of decoding correctly the information delivered by the source, as they could explore location diversity and correlate different measurements. Without any doubt, further investigations are needed to establish the robustness of current schemes against mobile UAV-equipped adversaries.}

\textcolor{black}{
\textbf{Untrusted UAVs Relays.} When UAVs are included in the scenario, all the works surveyed in this contribution considered trusted UAV relays, always cooperating with the intended source and destination. However, UAVs can be also hijacked by the adversary. Moreover, they could behave according to the \emph{honest-but-curious} paradigm, i.e., behaving as intended, but trying to access the information delivered from the source. While these scenarios could appear similar to the ones discussed in Section~\ref{sec:relay_adv}, the inherent mobility features of the UAVs can provide additional advantages to the adversary, as highlighted in the previous point. Hence, there is a need for further research in this challenging scenario, calling for solutions able to mitigate the presence of single/multiple untrusted UAVs relays.
}

\textbf{Active Adversaries.} The recent contribution by the authors in~\cite{do2019_sensors} has been the first to consider an active adversary, not only interesting in decoding the confidential information but also in the disruption of the main communication link. Indeed, a powerful adversary able to inject noise on the communication channel, eventually using a directional antenna pointed toward the target communication link, could decrease the SINR experienced by the legitimate destination, leading to an increased number of errors and increased guessing chances by the adversary itself. 

\textbf{Determination of Relay Trust.} The discussion in Section~\ref{sec:relay_adv} highlighted some directions to keep the information confidential from untrusted relays, at the cost of a destination-assisted or device-assisted jamming. However, the level of trust of the relays in the network should be established at the deployment or at the boot time, and not at run-time. This is crucial to avoid the unnecessary cost of friendly jamming in terms of network and interference management or to reduce network information exposure to a relay. Thus, dedicated mechanisms and strategies to determine the level of trust of a relay should be implemented. To the best of the authors' knowledge, these considerations have not been addressed in the contributions available in the literature.

\textcolor{black}{
\textbf{Clustering and Energy Availability Issues in EH-NOMA networks.} One of the key challenges when designing secure NOMA networks is the clustering logic. In this specific context, \emph{clusters} refer to groups of devices sharing the same frequency-time resource, while the clustering refers to the process of assigning resources to the clusters. We notice that optimizing the clustering of the users might involve undesired effects, such as the unwanted delivery of sufficient energy to eavesdroppers, enabling them to decode information and let the secrecy rate of the main communication links to fall below the desired threshold. 
Based on the above considerations, this is one of the key challenges to address when designing \emph{optimized} NOMA networks, especially for locations with medium-to-low SNR. The solutions could leverage mobile devices, such as UAVs, equipped with directional antennas, in a way to minimize the leakage of the signal to undesired/insecure locations.
}

\textcolor{black}{
\textbf{Energy Harvesting - Rate-Splitting Multiple Access (RSMA) schemes.} While NOMA schemes rely on the full decoding of the interference, Rate-Splitting Multiple Access (RSMA) schemes take into account the potentially destructive nature of the interference, and therefore, they treat the received signal partially as noise and partially as \emph{intended} constructive interference, decoding it~\cite{mao2019_icc}. 
The usage of RSMA schemes has been recently studied mainly in the context of cellular networks, in broadcast-channel single-cell scenarios~\cite{clerkx2020_wlett},~\cite{thomas2020_arxiv}, and interference-channel multi-cell setup~\cite{hao2017_tit}.  \\
Similar to the case of NOMA schemes, it is straightforward to apply RF EH principles with networks operating according to RSMA techniques. Indeed, hybrid access points could broadcast dedicated energy signals in the downlink to cellular and D2D users working according to the RF EH principles discussed in Sections~\ref{sec:wpcn} and~\ref{sec:swipt}.  Hybrid access points can further  gather the information transmitted by the users in the uplink, relaying them to the base station. In this context, several optimization problems can be formulated, in a way to reduce the energy consumption, minimize the delay in information reporting, or maximizing the throughput, eventually considering friendly jamming techniques and the imperfections/non-idealities of real EH circuits~\cite{mao2019_tcomm},~\cite{flores2020_arxiv},~\cite{li2020_arxiv}. \\
RSMA schemes have a high potential to be used for security applications. Indeed, RSMA schemes can split the user's message into a private part, decodable by only a specific user, using a fraction of the overall available power, and a common part, decodable by a given group of users, using the remaining part of the available power~\cite{hao2017_tit}. However, at the time of this writing, the literature does not provide any specific study investigating classical security-related metrics in the context of energy harvesting RSMA schemes, such as the secrecy rate, the secrecy outage probability, and the secrecy capacity. Note that this is a particularly challenging area, as the energy consumption of EH users should not exceed the related EH capabilities of the users. We expect this area to be particularly promising in terms of research contributions, in the months to come.
}

\textbf{Performance in Real Deployment Scenarios.} The recent work by the authors in~\cite{wang2019_infocom} has been the first to provide an experimental evaluation of the analytical and simulation findings in the RF EH physical-layer security area. Indeed, providing experimental testbeds and open-source code could help the research community to either validate and verify the findings of a specific work, as well as to easily extend that results in different scenarios. Moreover, the translation from the simulation to the experimental validation could help to refine the theoretical models and highlighting technical issues, such as network synchronization and interference cancellation. Furthermore, real experimental evaluations could help to translate from security evaluations based on SINR at the eavesdropper to the definition of \emph{secure deployment areas}, where the minimum required secrecy rate is achieved.
At this time, despite the presence of a large number of theoretical results validated by simulations, almost no experimental results or real testbeds (excluding the one mentioned before) are available in the literature.

\section{Additional Physical-Layer Countermeasures}
\label{sec:defense}

Techniques focusing on physical-layer data secrecy and lightweight cryptography schemes addressed only a limited part of the attacks and threats described in Section~\ref{sec:attacks}. The current literature includes also additional countermeasures, deployed on purpose to protect EH networks against specific threats. These approaches, falling neither in techniques aimed at protection data secrecy, nor in lightweight cryptography schemes, have been discussed in this section, and cross-compared along different common features. 

A security-oriented overview of such scientific contributions is provided in Table~\ref{tab:defense_eh}.
\begin{table*}[hbtp]
\centering
\caption{Overview of physical-layer protection techniques adopted in EH networks}
\begin{tabular}{|P{1.2cm}|P{2cm}|P{1.5cm}|P{2.5cm}|P{3cm}|P{2cm}|P{2cm}|}
\hline
\textbf{Scheme} & \textbf{Enabling Technologies} & \textbf{Scenario} & \textbf{Considered Threat} & \textbf{Defense Tools} & \textbf{Target Security Service} & \textbf{Analysis Tool} \\ \hline\hline
\cite{Shakhov2013} & Generic & WSN & Flooding Attacks & Throughput degradation estimation, Intrusion Detection Based, Path Reassignment, Traffic aggregation & Availability & Markov Process \\ 
\hline
\cite{belmega2017_tifs,belmega2017_icc} & Generic & Generic EH Wireless Network & Jamming & Channel Hopping (OFDM), Power Spreading & Availability & Game Theory \\ 
\hline
\cite{rezgui2019} & Generic & Generic EH Wireless Network & Jamming & Time Splitting EH & Availability & Game Theory \\ 
\hline
\cite{hoang2015} & Cognitive Radio Networks & Generic EH Wireless Network & Jamming & Deception Tactic (fake transmissions) & Availability & Markov Process \\ 
\hline
\cite{knorn2019_csl} & Generic & Wireless Communication Control Systems & Jamming & Optimal non-causal Energy Allocation Policy & Availability & Markov Process \\ 
\hline
\cite{guo2017_twc} & Interference Alignment Wireless Network & Generic EH Wireless Network & Jamming & Opportunistic Interference Alignment & Availability & MinIL Algorithm \\  
\hline
\cite{pu2017} & Generic & IoT, WSN & DoS (network level), Stealthy Collision Attacks & Adaptative Acknowledgement Approach (AAA) & Availability & Markov Process \\ 
\hline
\cite{chang2019_tons,chang2017_wisec}& Generic & Generic EH Wireless Network & Energy DoS & Power Positive Networking (PPN) & Availability & Resurrection Duckling Based \\ 
\hline
\cite{liu2016_commag} & WPCN & Generic EH Wireless Network & Jamming, Spoofing, Monitoring, Charging Attacks, Malicious Applications & FDMA, TDMA, frequency hopping, Interference Alignment; Digital Signatures; Data and Energy exchanging estimation & Availability, Confidentiality & Experimental \\ 
\hline
\cite{luo2018_sensys} & ABCN & IoT & Jamming, Deauthentication, Replay, Channel Spoofing & Ambient Backscattering, Multi-path Propagation Signatures & Availability, Authenticity & One-Class SVM \\ 
\hline
\cite{moukarzel2017} & Generic & Generic EH Wireless Network & Power and Timing Side Channels & Quantization Controllers & Confidentiality & Implementation Optimization \\ \hline
\cite{Yang2017_mobicom, Luo2018} & ABCN & Mobile Devices, WBAN & Side Channels, Authenticity & Shielding, Software Layer Solutions, Backscatter Signal Authentication with Propagation Signatures & Confidentiality, Authenticity & Physical-Layer Analysis \\ 
\hline 
\cite{krishnan2019} & Generic & Intermittent Computing Systems & Power Interruption Attacks & Secure Checkpoints & Confidentiality, Integrity, Authenticity, Availability & Input/Output \\ 
\hline
\end{tabular}
\label{tab:defense_eh}
\end{table*}

The following discussion provides more details about the schemes listed in Table~\ref{tab:defense_eh}, introducing them based on the target security service.

Many scientific contributions in the last years, such as \cite{belmega2017_tifs}---\nocite{belmega2017_icc,rezgui2019,hoang2015,knorn2019_csl}\cite{guo2017_twc}, provided physical-layer solutions to guarantee the availability of an EH network. More in detail, to face malicious jamming sources, the contributions in~\cite{belmega2017_tifs, belmega2017_icc, rezgui2019, guo2017_twc} recommended the use of well-known frequency hopping and spreading spectrum techniques. Other contributions, such as \cite{belmega2017_tifs, belmega2017_icc, rezgui2019} convert the jamming signal power into useful transmission power, used to charge the legitimate devices. Following game theory concepts, they modeled the interaction between a pair of legitimate nodes and a jammer as zero-sum-game, and they found optimal solutions (i.e., equilibrium conditions) that minimize the impact of the jammer and maximize the harvesting energy on the legitimate devices.

\noindent
In the context of Energy Harvested \acl{CRN}, the authors in~\cite{hoang2015} considered the use of a deception tactic to exhaust energy resources of the attackers, i.e., to send fake transmission data to compromise the attackers' capabilities, to deplete their energy. This countermeasure allows the secondary users in a cognitive radio network to deal with jamming attacks, reducing the impact of the attack. The authors in~\cite{knorn2019_csl} proposed a solution against jamming attacks in closed-loop systems, by working on the solution of the optimization energy allocation problem. An optimal non-causal energy allocation policy allows the system to estimate the state of the control process (leveraging a Kalman Filter) while ensuring availability and stability.
Similarly, the authors in~\cite{guo2017_twc} considered interference alignment methods (i.e., beamforming strategies) to mitigate adversarial jammers. The anti-jamming solution has the main goal of creating an Interference Alignment network comprising some of the legitimate devices (e.g., the transmitter and the \ac{EH} nodes), while the remaining nodes are dedicated to EH tasks. Both the interference signal and the jamming signal are considered as the sources of RF EH on the compatible devices.

\noindent
A protection scheme to guarantee network availability under flooding attacks in the WSN scenario has been proposed by the authors in~\cite{Shakhov2013}. Specifically, they used EH concepts and Markov Process theory to identify ongoing flooding attacks. They modeled the traffic state of the network via a Markov process, and they provide an estimation of the sensor node throughput degradation under flooding attacks, comparing it to a threshold. The higher the degradation, the higher the probability of ongoing flooding attacks. 

\noindent
To face DoS attacks, the authors in~\cite{pu2017} conceived a framework to identify stealthy and malicious nodes that transmit packets for intentional packet drop in a \ac{EH} network. They exploited the \ac{PDR} feature and an Adaptive Acknowledgment Approach (AAA) to identify these stealthy collision attacks. In the AAA strategy, each node sends a data packet, monitors the next transmission of its first hop downstream node and waits for the acknowledgment from the second hop downstream node. If the acknowledgment is not received in the timing window, there is a high probability of an attack.

\noindent
The authors in~\cite{chang2017_wisec} and~\cite{chang2019_tons} provided methods to mitigate DoS attacks, by offloading the power requirements of the EH node to the device that is sending networking requests. Specifically, they designed an alternative channel, namely Power-Positive-Networking (PPN), where each received signal allows to recharge the battery of the device. Being very similar to the underlying concept of SWIPT networks (see Section~\ref{sec:swipt}), these alternative channels stop energy DoS attacks, given that any jamming signal aimed at interrupting wireless charging results in a re-charge of the EH device.

\noindent
The authors in \cite{liu2016_commag} highlighted qualitatively some possible countermeasures that could be used in \ac{WPCN} against various types of attacks. To face jamming attacks, the authors proposed to leverage traditional modulation schemes, such as \ac{FDMA}, \ac{TDMA}, and frequency hopping, or to use interference alignment principles (beamforming). To thwart spoofing attacks, they indicated the possible use of digital signatures, to provide authenticity and integrity of the messages. To detect monitoring attacks, they proposed to listen to the communication channel and discover the presence of data exchanged between the energy harvester nodes and possible malicious nodes. Concerning charging attacks, the authors proposed to perform frequent estimations of the energy that nodes are harvesting in the network, and to detect anomalies in the charging profile. Finally, to defeat malicious applications, the authors proposed to verify the application signatures and to check if they are trusted/valid. Alternatively, they indicated the possibility to design energy-aware mechanisms, such that every application could leverage an energy budget to be used for computations and transmissions. Finally, to address the privacy issues discussed in Section~\ref{sec:attacks}, they proposed to collect the RF fingerprints from the environment, and to compare them with the ones acquired at run-time.

\noindent
In the context of the IoT networks powered via ambient backscatter, the authors in~\cite{luo2018_sensys} identified the possibility to enforce the authenticity of the communication via multi-path propagation signatures, thus securing pairing and data communication processes. They found that the interaction of the RF source with the environment and the location of the specific device is unique. Thus, it can be used to provide a probabilistic and ambient-dependent authentication, as well as a tool to identify replay and de-authentication deadlock attacks. 

\noindent
Other approaches investigated solutions to avoid side-channel attacks. 
For instance, the authors in~\cite{moukarzel2017} investigated the hardware design of a system adopting quantization controllers, using the harvested energy to carry out a secure computation in an isolated environment.

\noindent
In the ABCN scenario, the authors in~\cite{Yang2017_mobicom} proposed two strategies, i.e.: (i) to block the incident/reflected signal; and (ii) to adopt software solutions (e.g. set permissions on a device file, stop the network manager, hiding the WiFi interface). The usage of propagation signatures to identify the attacker has been also proposed and discussed by the authors in~\cite{Luo2018}.

\noindent 
Finally, to face \emph{power interruption attacks} at the physical layer, the authors in~\cite{krishnan2019} proposed a secure protocol for devices with intermittent connectivity, characterized by several security features. The protocol uses random nonces to protect against checkpoint replays attacks, leverages a pre-shared key to avoid sensitive data leakage, exploits hash chains to ensure data integrity, preserves the correct order of the operations by ensuring the consecution between the checkpoints, and, finally, improves the resilience of the device towards power loss issues by storing redundant copies of the system state.

\subsection{Lessons Learned}
\label{sec:defense_lesson}

The non-cryptographic protection strategies and tools discussed in the previous section highlighted several take-home messages and guidelines to be considered when deploying EH technologies. The key considerations are summarized below.


\textbf{Protecting the Availability of EH Networks.} Most of the solutions previously discussed focused on improving the resilience of the network against attacks targeting its availability. Indeed, differently from traditional IoT networks, energy provisioning in EH networks is often intermittent, and thus denying its availability could be the easiest way to disrupt network operation. While traditional means such as solar, wind, and mechanical sources are easiest to be disrupted, stopping RF harvesting is not easy. Hence, jamming the source only increases the amount of RF energy available, resulting in an increased amount of power available at the devices. This is the reason motivating the even more diffused consideration of the RF source as a secure harvesting method.

\textbf{Side-Channel Attacks in EH Networks.} Traditional EH sources, as well as RF harvesting solutions, have not been designed with physical security in mind. Therefore, well-known side-channel attacks are even more powerful and easy-to-perform than in traditional wireless networks. Due to the hard integration of classical hardware approaches, countermeasures available in the literature use physical-layer techniques (such as the identification of propagation signatures), whose security is probabilistic and not fully tested.

\subsection{Future Directions}
\label{sec:defense_future}

Few future research directions concerning lightweight non-cryptographic countermeasures for EH networks can be identified.

\textbf{Experimental energy expenditure evaluations.} Despite being effective, none of the above-described protection technologies provided an evaluation of the related energy consumption. This is a crucial aspect, because of the energy-security trade-off vastly described in this paper. Currently, it is still not clear if traffic analysis and physical-layer security techniques are effectively less energy-consuming than traditional cryptography approaches, especially considering the optimizations of cryptography techniques described in Section~\ref{sec:light_crypto}.

\textbf{Detecting Physical Presence in EH Networks.} As described in the previous subsection, a side-channel attack is still one of the most challenging threats in EH networks. It is worth noticing that side-channel attacks require physical adversaries, able to approach the target devices and deploy dedicated tools. Hence, to prevent such an attack, a solution to detect the presence of a physical intruder could be coupled with the deployment of EH networks, providing viable protection schemes. For instance, considering the sensitivity of RF EH technologies to modifications of the surrounding environment, physical adversaries can be identified, triggering appropriate response strategies. At the same time, further research directions could evaluate the feasibility of running such systems on the EH device or dedicated battery-supplied nodes. 

\section{Emerging Research Challenges}
\label{sec:challenge}
The previous sections highlighted the most prominent research areas involving EH devices and networks, as well as some future directions that can be pursued in the related macro-areas. Besides these promising study areas, there are a few other security application domains involving EH principles and devices that, over the last years, have received little (if any)  attention from the scientific community. Some of them are described below, along with key challenges to be solved to finally unlock their potentials.  

\textcolor{black}{
\textbf{Intelligent Surfaces.} Intelligent surfaces have recently gained momentum, as one of the most appealing new attractive technologies in the wireless communications research domain~\cite{gong2019_arxiv}. In the literature we often found an interchangeable use of the terms \ac{LIS} and \ac{IRS}. Overall, \ac{LIS} mainly refer to Active LIS-based Massive MIMO, as discussed in~\cite{hu2018_tsp}. They are used for data transmission (while IRS are used for reflection), and they consist of massive antenna arrays, organized on surfaces, and characterized by a fixed transmit power per volume-unit constraint. At this time, they are mainly considered for indoor applications, where they are shown to be particularly successful for interference cancellation~\cite{nadeem2019_arxiv}.
\\
Without loss of generality, an \ac{IRS} can be defined as a bidimensional planar surface, consisting of a large number of cheap and passive electromagnetic elements, structured in arrays. The phase shift of the reflecting elements in the array can be controlled independently via software, without requiring a dedicated energy source, in a way to reflect the incident signal with specific desired properties, such as attenuation and scattering~\cite{wu2020_commag}. Therefore, any software-controlled change in the coefficients of the IRS contributes to shape the incident signal with specific amplitude and phase, modifying the wireless propagation environment and boosting the reception of the signal towards specific locations~\cite{li2020_wcomlett}. Compared to existing solutions, such as ABCNs, Amplify-and-Forward schemes, and LISs, IRSs offer real-time flexibility and tuning of the coefficients, without requiring any additional energy budget or additional information to be delivered on the wireless channel~\cite{chen2019_access}.
\\
From the security perspective, intelligent surfaces have already started to attract the interest of many researchers, especially the ones working on physical-layer security and data secrecy schemes, as thoroughly discussed in Section~\ref{sec:phy_secrecy}. While LIS have been mainly investigated from the performance perspective, many very recent contributions have focused on the security properties offered by IRS.
Indeed, using IRSs it is possible to precisely control the phase, the amplitude, the frequency, and the polarization of the signals, including neither complex encoding/decoding schemes, nor energy-consuming processing operations. The usage of IRSs is particularly appealing when the potential eavesdroppers are closer to the source than the destination, or when the adversaries lie in the same direction of the intended destination of the message~\cite{wu2020_commag}.
In addition, the number of reflecting elements, as well as the specific phase shift adopted by the reflecting elements, and the IRS's reflect beamforming, can be all combined and optimized to restrict the received beams to specific locations, to cancel beams at the location of the eavesdroppers, to reduce the energy consumption on the transmitter, or to maximize the SINR at the intended receivers~\cite{basar2019_access}. \\
Therefore, research groups working on physical-layer security have already started to consider schemes that can optimize one or more metric, by taking into account the remaining system or security-related metrics as constraints of the optimization problems. In this direction, we can notice that the literature already includes some works that apply the findings and results discussed in Section~\ref{sec:phy_secrecy} to the new scenario, considering the configuration of the coefficient of the IRS as new parameters of the optimization problems. \\
For instance, the authors in~\cite{makarfi2019_arxiv},~\cite{feng2019_arxiv},~\cite{feng2019_iwcsp},~\cite{cui2019_wlett} have already provided preliminary studies considering single external eavesdroppers, equipped with single receiving antennas. Internal adversaries, in the context of Cognitive Radio Networks (CRN), have been already considered by the authors in~\cite{chen2019_access}. Single eavesdroppers equipped with multiple antennas have been considered in~\cite{jiang2020_arxiv}, while other very recent contributions considered non-colluding multiple eavesdroppers, equipped with either single~\cite{lu2020_arxiv} or multiple antennas~\cite{hong2020_arxiv},~\cite{yu2019_arxiv_2}. Other recent works, such as~\cite{yang2020_arxiv} are also evaluating the benefits of Artificial Intelligence (AI) techniques to boost the secrecy rate towards multiple eavesdroppers. Similarly to the studies discussed in Section~\ref{sec:phy_secrecy}, we also notice different formulations of the problem, based on the availability (or not) of the CSI~\cite{yu2019_arxiv}. 
In line with the previous discussion, the objectives pursued by the proposed optimization problems are very similar to the ones discussed in Section~\ref{sec:phy_secrecy}, such as the Secrecy Rate Maximization~\cite{shen2019_commlett} and the definition of the secrecy capacity and outage probability~\cite{makarfi2019_arxiv}, while other works consider the secrecy rate as a constraint of the optimization problem~\cite{chu2020_wlett}. Recent works are also considering the joint effect of friendly jamming, as a way to further improve the secrecy of the main communication links against external adversaries~\cite{guan2020_wlett},~\cite{xu2019_arxiv}.\\
Overall, following the current trend, we expect that a significant number of scientific contributions will appear in the upcoming months in the IRS area. Indeed, all the studies and findings consolidated in the last years in the context of data secrecy in EH networks can be revisited in the emerging IRS context, taking into account also the new degree of freedom offered by the coefficients of the IRS and their energy efficiency, paving the way for new appealing research challenges. For instance, to the best of our knowledge, none of the contributions in the current literature considered the chances of multiple colluding eavesdroppers to reduce the secrecy rate of networks using IRS, as well as none of them included considerations on the particular directivity of the antennas used by the legitimate devices and the eavesdroppers. In addition, new security challenges emerge: an attacker can take the control of the IRS, and modify at will the pattern of the incident signals (e.g., directing the beam towards unintended locations, at the advantage of the attacker); moreover, an attacker can also locate the IRS and get physically closer, improving its SINR.
We expect also that the introduction of IRS will move data secrecy and physical-layer security issues to new application domains, such as vehicular networks, Flying-Ad-hoc Networks (FANET), and Ultra-Dense Networks (UDN).
}

\textbf{Physical-Layer Authentication using EH Signals.} Some recent contributions, such as~\cite{rahman2015}, \cite{Luo2018}, and~\cite{xu2019_tmc}, investigated physical-layer authentication schemes leveraging EH properties and devices. However, the authors in~\cite{rahman2015} and~\cite{Luo2018} used physical-layer features, i.e., the frequency offset and the on-body backscattering reflections related to wave propagations, respectively, and contextualized their usage when EH devices are involved, thus not fully exploring EH features. Instead, the authors in~\cite{xu2019_tmc} were the first and only to use properties specific to EH phenomena, i.e., the profile of the energy harvested by a human during his regular gait. Using the EH profile over time derived by EH processes, they created a gait-based authentication protocol that does not require the continuous sample of the accelerometer, drastically reducing the overall energy consumption (up to $84$\%). In this context, the authors reported slightly lower performance than typical gait-based authentication schemes ($89$\% in harsh conditions) and significant success performance in spoofing attacks (up to $14.1$ \%). Thus, further enhancements are possible, e.g., by coupling the gait with the movements of other parts of the human body (e.g., hands and wrists). Besides, the methodology could be even re-used in other contexts, such as the authentication of embedded devices in industrial plants, that can be characterized by distinctive features.  

Overall, we can notice that despite physical-layer authentication schemes have been investigated in the context of RFID (see, for instance, \cite{yao2019_access}) and traditional backscattering systems (e.g., the works in ~\cite{Ranganathan2015} and~\cite{Yao2016}), the usage of EH properties for devices' authentication still lacks complete studies. For instance, studies about the unique charging profile of EH devices or the presence of any imperfection in the EH transfer are still not available in the literature, and can pave the way to a completely new wave of physical-layer authentication protocols, possibly overcoming the limitations of current schemes.

\textcolor{black}{
\textbf{Integrated Space and Terrestrial Networks (ISTN).} In their intended design, 5G wireless systems are mainly terrestrial networks, and therefore their coverage area is limited. 6G systems are expected to bring Low-Earth-Orbit (LEO) satellites into the system architecture, to provide worldwide broadband connectivity, including users in open sea and remote areas~\cite{rajatheva2020_6gwp_arxiv},~\cite{oligeri2020_wisec}. Despite LEO satellites are the most feasible to accomplish this task, due to the reduced distance from the Earth and the consequently reduced latency, the prevalence of the Line of Sight (LOS) component creates co-channel interference effects, and in turn, signal degradation effects. From the security perspective, LEO satellites are far from the ground, and very difficult to be hijacked physically. However, due to their distance, they can hardly be modified ex-post in their hardware features, or augmented with new enabling technologies. Therefore, their security properties should be carefully verified and tested before deployment. In this context, harvesting technologies can play a dual role. On the one hand, solar cells integrated into satellites can boost the energy available on the satellites, and increase the power available for transmission and relay operations. On the other hand, physical-layer information-theoretic security techniques can be combined with dedicated harvesting technologies leveraging Free Space Optical (FSO) communication links to further enhance the security of the communication links and, at the same time, increase the energy available at the receivers~\cite{li2020_iotj}.
}

\textbf{Physical Unclonable Functions using EH Principles.} In line with previous considerations on authentication using EH principles, recently the authors in~\cite{nozaki2018_imfedk} described how to use the dispersion of the power generation time of the solar cells to identify the specific semiconductor, and thus the device using the generated power. The usage of similar principles in other harvesting technologies, such as RF, still has not been investigated in the literature, and can provide anti-counterfeiting strategies, hardware random number generation tools, and auxiliary authentication solutions for EH devices.  

\textbf{Key Agreement Schemes using EH Properties.} Despite a large number of scientific contributions investigated innovative and lightweight key agreement schemes using physical-layer security mechanisms (see~\cite{mukh2014} for an overview), key establishment mechanisms in EH networks have received only minimal attention. A pioneering contribution in the context of \ac{ABCN} is the one described in~\cite{lv2016_icc}, where the authors got rid of the secret reconciliation procedure typical of physical-layer key agreement schemes, improving the efficiency of such schemes. However, the usage of physical-layer features specific to energy harvesting, such as the ratio of harvested energy over time and its relationship with unique devices' features for key agreement, still has not received enough attention by the scientific community.

\textbf{Privacy Considerations.} EH devices, especially IoT devices using EH capabilities, could be subject to potential privacy breaches, derived by the analysis of the energy consumption and re-charging patterns, as well as their positioning toward the source. Few recent scientific contributions, such as the ones by the authors in~\cite{giaconi2018_tifs} and~\cite{min2019_iotj}, highlighted these privacy issues in electricity and healthcare scenarios, respectively. At the same time, the authors in~\cite{tan2013_jsac} identified the potential of RF harvesting properties for enhancing the privacy level of smart meters, as they can differentiate energy sources and decorrelate the final consumed energy from the profile of the energy expected from the legitimate source, thus defending against side-channel attacks. The challenge here is to come up with privacy-preserving frameworks, that can mask the re-charging profiles and the energy expenditure behavior, while minimally affecting the system availability (i.e., being characterized by reduced energy consumption). 

\textbf{Secure Protocol Stack for EH Devices.} The unique systemic features of EH devices and networks, highlighted in the previous sections, require the design of dedicated solutions at all the layers of the protocol stack. Besides the significant work on physical layer security described in previous sections, several contributions investigating MAC protocols tailored for EH devices have been classified and described by the authors in~\cite{sherazi2018_adhoc}. At the same time, few contributions such as~\cite{Alrajeh2013} and~\cite{tang2018} investigated secure routing protocols for EH networks, mainly concentrating on reducing the energy footprint of such schemes. Especially when the EH network is targeted by attacks disrupting its availability, dedicated routing protocols are required, that
can rebalance the network traffic towards nodes with high energy-budget, thus improving its overall throughput. 

\textbf{Using Energy Harvesting for Security-Related Applications.} Finally, only a few contributions are describing commercial products using EH principles to enhance the physical security of devices and people. For instance, the authors in~\cite{zungeru2019_access} recently presented the implementation of a physical access control system based on energy harvesting, regulating the access of people to rooms in a building. However, several additional applications are possible, not only limited to the smart home application scenario, but also extending to industrial control systems, entertainment, and intelligent transportation (e.g. autonomous vehicles, UAVs), to name a few.

\section{Conclusion}
\label{sec:conclusion}

In this paper, we survey the prominent security issues, mechanisms, techniques, and applications related to \acl{EH} for wireless networks. 
In particular, we first provide background information on models and technologies necessary to delve into the EH field. Later, we address the security domain. In detail, after a discussion of the possible attacks, we have partaken the security domain for EH wireless networks into three main macro-areas, i.e.: data secrecy at the physical layer; lightweight cryptography techniques; and, additional physical-layer countermeasures. Within these areas, we classified the available scientific contributions along some common features, such as the assumed adversarial model and the target security service, we discussed common approaches and distinguishing strategies, and we identified future promising research directions. Furthermore, we have highlighted appealing research problems related to EH networks, such as Intelligent Reconfigurable Surfaces (IRS) and physical-layer security issues in Rate-Splitting Multiple Access schemes.
We believe that the exposed challenges show that the development of security techniques for EH networks is still an exciting research area, and that it can be inspiring for researchers, industry, and start-ups, striving for innovative, non-invasive, and computationally lightweight means to enforce systems security in constrained energy-limited environments.

\section*{Acknowledgements}
The authors would like to thank the anonymous reviewers for their comments and suggestions, that helped improving the quality of the manuscript.\\*
This publication was partially supported by awards NPRP11S-0109-180242 and NPRP X-063-1- 014 from the QNRF-Qatar National Research Fund, a member of The Qatar Foundation. The information and views set out in this publication are those of the authors and do not necessarily reflect the official opinion of the QNRF.

\bibliographystyle{IEEEtran}
\bibliography{eh}

\begin{IEEEbiography}[{\includegraphics[width=1in,height=1.15in,clip,keepaspectratio]{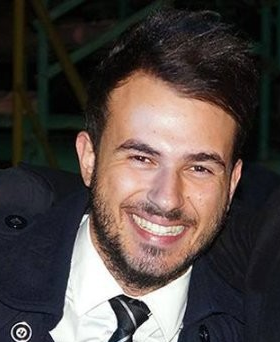}}]{Pietro Tedeschi} is a PhD Student in Computer Science and Engineering at HBKU-CSE, Doha, Qatar. He is an active member of the Cyber-Security Research Innovation Lab. He received his Bachelor's degree in Computer and Automation Engineering in 2014 with a thesis on the Analysis of Security Protocols for the Internet of Things, in IEEE 802.15.4e Networks, and his Master's degree (with honors) in Computer Engineering in 2017 with a thesis on the Development of Security Architectures in Intelligent Transport Systems for EU Horizon 2020 BONVOYAGE project, both from the ``Politecnico di Bari``. From 2017 to 2018, he worked as Security Researcher at CNIT (Consorzio Nazionale Interuniversitario per le Telecomunicazioni), Italy, for the EU H2020 SymbIoTe project. His research interests span over UAV/Drone Security, Wireless Security, Internet of Things (IoT), Applied Cryptography, and Cyber-Physical Systems. 
\end{IEEEbiography}
\begin{IEEEbiography}[{\includegraphics[width=1in,height=1.15in,clip,keepaspectratio]{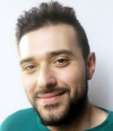}}]{Savio Sciancalepore}
Savio Sciancalepore is currently Post Doc at HBKU-CSE-ICT, Doha, Qatar. He obtained his Master's degree in Telecommunications Engineering in 2013 and the PhD in Electric and Information Engineering in 2017, both from the Politecnico di Bari, Italy. He received the prestigious award from the ERCIM Security, Trust, and Management (STM) Working Group for the best Ph.D. Thesis in Information and Network Security in 2018. He is also serving as an Academic Editor for the journal Security and Communication Networks (Hindawi). His major research interests include network security issues in Internet of Things (IoT) and Cyber-Physical Systems.
\end{IEEEbiography}
\begin{IEEEbiography}[{\includegraphics[width=1in,height=1.15in,clip,keepaspectratio]{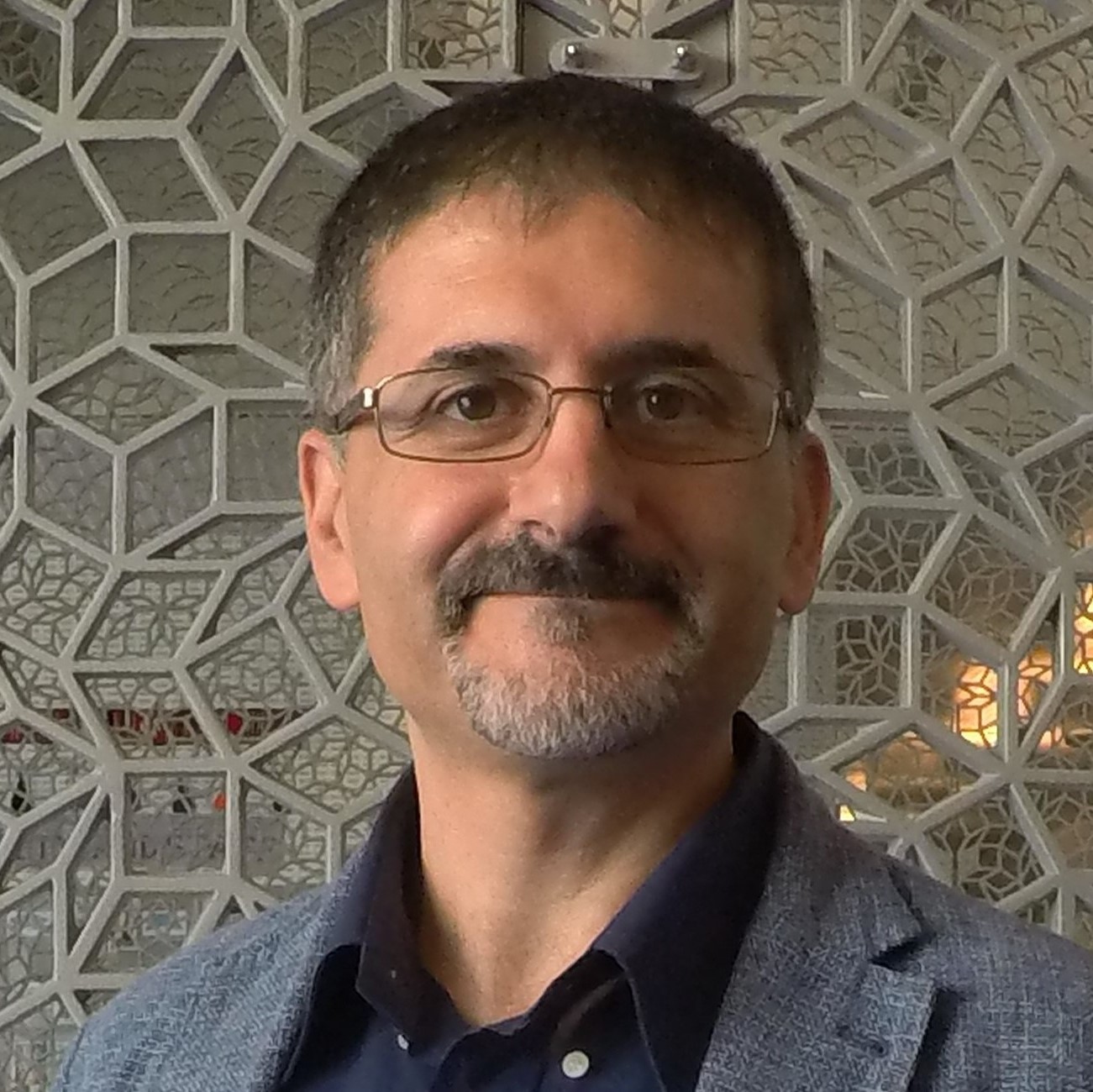}}]{Roberto Di Pietro,} ACM Distinguished Scientist, is Full Professor in Cybersecurity at HBKU-CSE. Previously, he was in the capacity of Global Head Security Research at Nokia Bell Labs, and Associate Professor (with tenure) of Computer Science at University of Padova, Italy.  He also served 10+ years as senior military technical officer. Overall, he has been working in the cybersecurity field for 23+ years, leading both technology-oriented and research-focused teams in the private sector, government, and academia (MoD, United Nations HQ, EUROJUST, IAEA, WIPO). His main research interests include security and privacy for wired and wireless distributed systems (e.g. Blockchain technology, Cloud, IoT, On-line Social Networks), virtualization security, applied cryptography, computer forensics, and data science. 
Other than being involved in M\&A of start-up---and having founded one (exited)---, he has been producing 230+ scientific papers and patents over the cited topics, has co-authored two books, edited one, and contributed to a few others. 
He is serving as an AE for ComCom, ComNet, PerCom, Journal of Computer Security, and other Intl. journals. In 2011-2012 he was awarded a Chair of Excellence from University Carlos III, Madrid. In 2020 he received the Jean-Claude Laprie Award for having significantly influenced the theory and practice of Dependable Computing. 
\end{IEEEbiography}

\end{document}